\documentclass{ametsocV6.1}
\usepackage{hyperref}
\usepackage{xr}
\usepackage[english]{babel}
\usepackage[autostyle, english = american]{csquotes}
\MakeOuterQuote{"}
\makeatletter
\nolinenumbers
\newcommand*{\addFileLabels}[1]{%
  \edef\sigma@temp{#1}%
  \expandafter\input\expandafter{\sigma@temp.aux}%
}
\makeatother
\title{Moisture Budgets and Circulation Analogs: Diagnosing Dynamic and Thermodynamic Precipitation Change}

\authors{Robert Doane-Solomon\aff{a}\correspondingauthor{Robert Doane-Solomon, robert.doane-solomon@physics.ox.ac.uk}\\
Isla R. Simpson\aff{b}\\
Tim Woollings\aff{a}}

\affiliation{\aff{a}{Atmospheric, Oceanic and Planetary Physics, University of Oxford, Oxford, UK}\\
\aff{b}{Climate and Global Dynamics Laboratory, NSF National Center for Atmospheric Research, Boulder, CO, USA}}

\abstract{
Precipitation trends can arise from both dynamic factors (changes in atmospheric circulation) and thermodynamic factors (changes in atmospheric moisture content). Disentangling these contributions can aid in understanding regional climate change and improving projections. We compare two approaches which separate dynamic and thermodynamic contributions to precipitation trends over Central Chile: a moisture budget analysis and constructed circulation analogs. Both methods are applied to fields from the CESM2 Large Ensemble as well as two reanalyses. We analyze the methodological differences that lead to distinct results in each approach and evaluate their respective capabilities in capturing dynamic, thermodynamic and coupled trends. 
We find that the estimated dynamic trends from both methods often differ substantially for individual ensemble members, although the ensemble mean generally agrees in sign but not in magnitude. Finally, we apply circulation analogs to moisture budget terms to refine estimates of historical and future precipitation change in Central Chile. This combined framework reduces dynamic contamination of thermodynamic trends and yields projections of thermodynamic precipitation change that are weaker than that suggested by either method alone.}

\begin{document}

\maketitle


%
%
\statement
	In order to better understand changes in regional precipitation, we need to understand which physical mechanisms are behind the change in precipitation. There are two main mechanisms: precipitation can change simply because the atmosphere is getting warmer, or because the atmospheric circulation is changing as well. Previous studies have used multiple different methods to understand which physical processes are dominant in different regions, but no method is perfect. In this paper we compare and combine two methods in a case study region of Central Chile, to better understand the methods themselves and help refine our understanding of why rainfall is declining in this region.

\section{Introduction}
Precipitation changes are projected to occur worldwide as a response to global warming \citep{DORE20051167, held_robust_2006,seager_thermodynamic_2010}. In some locations, such as Mediterranean-type climate regions, precipitation declines have already occurred and are projected to continue through the 21st century \citep{polade_precipitation_2017, seager_climate_2019,seager_recent_2024,doane2025dynamic}. Located at the subtropical western edge of continents, these regions are underneath semi-permanent subtropical highs in the summer linked to the descending branch of the Hadley cell. In winter, the Hadley cell retreats towards the opposing hemisphere, allowing the extratropical storm track to deliver rainfall to these regions. The majority of their precipitation occurs in wintertime from these systems \citep{seager_climate_2019,doane2025dynamic}, making them especially susceptible to hydroclimate change because most of their annual rainfall is concentrated in a single season.

There are two primary mechanisms which can cause precipitation change: dynamic and thermodynamic. Dynamic precipitation changes are caused by changes in circulation. Examples of this could be the strengthening Southern Hemisphere storm track \cite{o2010understanding, chemke2022intensification, kang2024revisiting}, or the poleward movement of the Hadley cell edge leading to an expansion of the subtropical dry zones \citep{lu2007expansion, grise2019recent,chemke2023human}. Thermodynamic changes occur because warming increases the saturation vapour pressure, enabling the atmosphere to hold more water vapour, thereby altering moisture gradients and relative humidity even in the absence of circulation changes. This dependence is described by the Clausius Clapeyron equation, which gives the temperature sensitivity of the saturation vapour pressure $e_s$ of one mole of water:

\begin{equation}
\frac{de_s}{dT} = \frac{L}{T \Delta V},
\label{cc-raw}
\end{equation}

where $L$ is the latent heat of vaporisation and $\Delta V = V_{vapor} - V_{liquid}$ is the molar volume change between vapour and liquid. On this basis, \cite{held_robust_2006} argued that relative changes in precipitation minus evaporation, $\frac{\delta(P-E)}{P-E}$, should scale with temperature at roughly 7\% per Kelvin. In practice, the response deviates from this simple scaling due to the effect of warming on relative humidity and temperature gradients, among other factors \citep{byrne_response_2015,siler2023diagnosing, braschoss2025state}.

One can use a continuity equation for conservation of moisture to separate precipitation changes into dynamic and thermodynamic processes. The specific humidity $q$ within an atmospheric column, $\int_{p=0}^{p=p_s} q \, dp$, can change due to advection and convergence by the horizontal velocity $\textbf{u}$, as well as precipitation $P$ and evaporation $E$. Precipitation acts as sink for water vapor, and evaporation acts as a source. To ensure that $P$ and $E$ are in units of meters per second (or mm/day), we introduce the constants $g$, the gravitational field strength, and $\rho_{w}$, the density of water \citep{seager2013diagnostic}. This leads to the following continuity equation:

\begin{equation}
    \frac{1}{g\rho_{w}} \left[ \frac{d}{dt} \int_0^{p_s} q \, dp + \nabla \cdot \left(\int_0^{p_s} q \mathbf{u} \,\, dp \right)\right] = E - P.
    \label{continuity}
\end{equation}

\noindent When considering timescales such as climatological monthly means, the time derivative term is negligible, so we can simplify Eq. ~\ref{continuity} to:

\begin{equation}
    P - E = -\frac{1}{g\rho_{w}} \nabla \cdot \left(\int_0^{p_s} q \mathbf{u} \,\, dp \right)
  = \text{VIMC},
    \label{continuity-simplified} 
\end{equation}

where VIMC is a shorthand for the vertically integrated moisture flux convergence. VIMC can then be decomposed into contributions from changes in $q$ (thermodynamic) and $\mathbf{u}$ (dynamic), although this separation is not exact because variations in $q$ may arise from both thermodynamic and dynamic processes. Nevertheless, moisture budget techniques based on this continuity equation have been widely used to understand the physical processes driving precipitation change \citep{seager_thermodynamic_2010, seager2014causes, seager2014dynamical, seager_climate_2019,  seager2013diagnostic, wills_stationary_2015, siler2023diagnosing, braschoss2025state}. 

An alternative framework for separating dynamic and thermodynamic influences is dynamical adjustment. Rather than decomposing the moisture continuity equation, these methods quantify the extent to which variability and trends in a target field, such as precipitation or VIMC, can be attributed to large-scale atmospheric circulation \citep{smoliak2015dynamical, deser_forced_2016, lehner_attributing_2018}. The underlying assumption is that the variability of the target field is strongly constrained by these circulation patterns. There may be other contributions to its variability, such as localized convection, soil moisture feedbacks, or changes in atmospheric humidity. By construction, these methods partition the target variable anomaly into a circulation-driven (dynamic) component and a residual component that captures all other processes, including thermodynamic changes.

A range of dynamical adjustment techniques have been developed \citep{sippel_uncovering_2019, pfleiderer_evaluation_2026}, including analog-based approaches such as constructed circulation analogs (CCAs). These have been used to dissect European and Mediterranean-type climate precipitation trends \citep{deser_range_2023, doane2025dynamic} as well as wider Northern Hemisphere precipitation changes \citep{guo_human_2019}. 

Despite the multiple studies using moisture budget diagnostics and dynamical adjustment techniques, a direct comparison of these two approaches applied consistently to the same region and variable has not yet been performed. Both methods provide useful but inherently imperfect decompositions of hydroclimate change, as each contains assumptions that limit a clean separation of dynamic and thermodynamic contributions. Here, we apply the constructed circulation analog dynamical adjustment technique and a Reynolds decomposition moisture budget analysis to VIMC in the Mediterranean-type climate region of Central Chile. This region is selected as a case study due to its clear declining precipitation trends in both observation-based and model datasets \citep{boisier_anthropogenic_2016, boisier_anthropogenic_2018, garreaud_central_2020, doane2025dynamic}. By comparing these methods, we assess how each partitions precipitation changes into dynamic and thermodynamic components, and evaluate whether they produce consistent physical interpretations of hydroclimate trends.

\section{Methods}
\label{methods}
\subsection{Data Sources}
We use precipitation ($P$), evaporation ($E$), specific humidity ($q$), zonal ($u$) and meridional ($v$) wind, surface pressure ($p_s$), and mean sea level pressure (SLP) from the ERA5 reanalysis \citep{era5pressurelevels, hersbach2023era5} and the JRA-3Q reanalysis \citep{ncar_gdex_dataset_d640000}. To calculate moisture budget and VIMC quantities, we use 6-hourly data of $q, u, v$ and $p_s$, and for circulation analogs we additionally use monthly mean values of SLP, $P$ and $E$. The monthly averages of 6-hourly moisture budget data are compared to post-processed VIMC data for ERA5 \citep{Mayer_ERA5_Budgets_2021}, which use mass-consistent horizontal wind fields and are only available as monthly averages (here referred to as PP-VIMC) to investigate the influence of the post-processing step. We also compute VIMC-monthly, which is simply the VIMC computed with monthly average $q, u, v$ and $p_s$: note that this is not equal to the monthly average of the VIMC computed at 6-hourly timesteps. 

For model data, we use a 100-member large ensemble  from the Community Earth System Model version 2 (CESM2-LE) \citep{rodgers2021ubiquity}, with nominal 1.0\textdegree\ horizontal resolution and monthly output. The ensemble consists of 50 members forced with CMIP6 historical forcing and 50 members forced with smoothed biomass burning from initialization in 1850 through 2014, followed by the SSP370 scenario from 2015 to 2100. No significant differences in trends were found between the two historical forcing configurations, so they are combined in this study. In addition, 10 members from the CMIP6-forced ensemble provide 6-hourly output, which we use to quantify submonthly variability. 

We also use a 2000-year preindustrial control simulation from CESM2, hereafter referred to as piControl. This simulation \citep{danabasoglu2020community} is a CESM2.1 coupled control simulation as part of CMIP6. It also has nominal 1.0\textdegree\ horizontal resolution and monthly output. To estimate trends from the piControl simulation associated internal variability, we construct overlapping segments such that no more than half of each period overlaps with any other. For example, 42-year trends are calculated over years 1–43, 22–64, 44–86, and so on. This yields 92 independent 42-year trends and 32 independent 120-year trends.

All reanalysis moisture budget quantities are calculated on a 1\textdegree \ $\times$  1\textdegree \ grid due to this being similar to the native resolution of CESM2. CESM2 moisture budget quantities are calculated on its native grid. All model and reanalysis fields are then conservatively remapped to a 0.5\textdegree\ $\times$ 0.5\textdegree\ to better delineate the coastal regions for masking and regional averaging. We apply an ERA5 land–sea mask on the same grid when computing precipitation and VIMC averages, using a 50\% land fraction threshold. 

We use the Mediterranean-type region of Central Chile as our case study: its boundaries are 30-42\textdegree S, 70-74\textdegree W. In the Supplementary Information we also use South Australia (30-39 \textdegree S, 114-142\textdegree E). We only analyze the Southern Hemisphere Mediterranean-type climate wet season MJJA. The regions used for the pressure analogs are the same as in \cite{doane2025dynamic}. All available pressure levels are used for moisture budget computation (37 for ERA5, 45 for JRA-3Q, 32 for CESM2).

We use reanalyses to estimate observed trends over the satellite era from 1980 to 2022. The CESM2 large ensemble is also used to increase the sample size over the same period and to assess whether differences arise between moisture budget and constructed circulation analog diagnostics. The piControl simulation is used to assess the extent to which apparent thermodynamic trends can arise in the absence of external thermodynamic forcing, and to evaluate whether recent dynamic and thermodynamic trends have emerged from the model’s internal variability. We consider two periods with the large ensemble and piControl, 1980-2022 (to compare with reanalyses) and also 1980-2100. All trends are calculated using OLS regression.  

\subsection{Moisture Budget Framework}
The moisture budget approach adopted here is a form of Reynolds decomposition applied to Eq.~\eqref{continuity-simplified}, similar to \cite{seager2013diagnostic}, \cite{seager_climate_2019} and \cite {tootoonchi_revisiting_2025}. 
We decompose specific humidity $q$ into $q = \bar{\bar{q}} + \bar{q}' + q''$ and $\mathbf{u}$ into $\mathbf{u} = \bar{\bar{\mathbf{u}}} + \bar{\mathbf{u}}' + \mathbf{u}''$, where the double overbar $\bar{\bar{()}}$ denotes the climatological mean over the full period analyzed, the single overbar and prime $\bar{()}'$ denote the monthly mean anomaly from the climatological mean, and the double prime $()''$ denotes the 6-hourly anomaly from the monthly mean. Expanding out the decomposition, VIMC can be written as follows:

\begin{equation}
-\nabla \cdot \int_0^{p_s} q \mathbf{u} \, dp =
-\nabla \cdot \int_0^{p_s} \Big[
      \bar{\bar{q}}\,\bar{\bar{\mathbf{u}}}
    + \bar{\bar{q}}\,\bar{\mathbf{u}}'
    + \bar{\bar{q}}\,\mathbf{u}''
    + \bar{q}'\,\bar{\bar{\mathbf{u}}}
    + \bar{q}'\,\bar{\mathbf{u}}'
    + \bar{q}'\,\mathbf{u}''
    + q''\,\bar{\bar{\mathbf{u}}}
    + q''\,\bar{\mathbf{u}}'
    + q''\,\mathbf{u}''
    \Big] dp.
    \label{expansion}
\end{equation}

Each of the nine terms represents a contribution of mean, monthly, and submonthly variations of $q$ and $\mathbf{u}$ and their interactions to the total vertically integrated moisture flux convergence. By construction, five of the nine terms have a constant monthly mean: $\bar{\bar{q}}\,\bar{\bar{\mathbf{u}}}$, $\bar{\bar{q}}\,\mathbf{u}''$, $\bar{q}'\,\mathbf{u}''$, $q''\,\bar{\bar{\mathbf{u}}}$ and $q''\,\bar{\mathbf{u}}'$. Thus, trends in these five terms can only result from trends in surface pressure $p_s$, which are small compared to the trends observed in the other four terms (not shown). These four remaining terms each represent a different physical mechanism that affects moisture convergence:

\begin{equation}
  -\nabla \cdot \int_0^{p_s} q \mathbf{u} \, dp \approx
  -\nabla \cdot \int_0^{p_s} \Big[
      \bar{\bar{q}}\,\bar{\mathbf{u}}'
      + \bar{q}'\,\bar{\bar{\mathbf{u}}}
      + \bar{q}'\,\bar{\mathbf{u}}'
      + q''\,\mathbf{u}''
      \Big] dp.
      \label{expansion_short}
  \end{equation}

\begin{itemize}
    \item $\bar{\bar{q}}\,\bar{\mathbf{u}}'$:
    This term represents changes in moisture convergence due to changes in the monthly mean wind, acting on the climatological mean moisture field.

    \item $\bar{q}'\,\bar{\bar{\mathbf{u}}}$:
    This term represents changes in moisture convergence due to changes in monthly mean moisture, acting through the climatological mean wind field.

    \item $\bar{q}'\,\bar{\mathbf{u}}'$:
    This term represents the contribution from the covariance of monthly mean moisture and winds to the total moisture convergence.

    \item $q''\,\mathbf{u}''$:
    This term represents the contribution from submonthly (e.g., synoptic-scale, transient eddy) covariability of moisture and winds to the total moisture convergence.

\end{itemize}

The total trend in VIMC can then be partitioned into the sum of these four contributions. Note that, due to Leibniz's rule for differentiation of an integral:

\begin{equation}
    \nabla \cdot \left(\int_0^{p_s} q \mathbf{u} \ dp\right)
  = \int_0^{p_s} \nabla \cdot (q \mathbf{u}) \, dp \ + \ q_s u_s \nabla p_s
    \label{leibniz} 
\end{equation}

The latter surface term may be quite significant \citep{seager2013diagnostic, seager_climate_2019}. However, it is not necessary to consider unless one also splits up $\nabla \cdot (q\mathbf{u})$ into its divergence $q\nabla\cdot\mathbf{u}$ and $\textbf{u} \cdot \nabla q$ advection components, which we do not do in this study. Therefore we keep VIMC in the form where the divergence operator acts after the vertical integral. In the rest of this manuscript, the vertical integral symbol is omitted for brevity.

\subsection{Constructed Circulation Analogs Framework}

Constructed circulation analogs belong to a broader class of methods known as dynamical adjustment, which aim to isolate the influence of atmospheric circulation on a target variable such as precipitation or temperature from other influences, often interpreted as thermodynamic. These methods decompose a target timeseries into a dynamically-driven component, representing the portion explained by circulation variability, and a residual component, which contains variability not explained by the circulation. The residual component may include thermodynamic effects as well as any  dynamical contributions not explained by the circulation analogs. We use a constructed circulation analog method nearly identical to that in \cite{doane2025dynamic}, which is similar to that in \cite{deser_forced_2016} and \cite{lehner_attributing_2018}. The method is as follows.

Using MJJA months only, we first compute monthly anomalies of SLP and our target field relative to the 1980-2022 climatology, or, if using piControl, the 43-year climatology from the selected period. For each target month in this period, we identify the $N_a$ MJJA months with the most similar SLP anomaly patterns, measured using a cosine-latitude-weighted Euclidean distance. To reduce sensitivity to individual analog choices, a random subset of size $N_s$ is drawn without replacement from these $N_a$ candidate months.

The target month SLP anomaly field is then expressed as a linear combination of the $N_s$ selected analog pressure fields through a multiple linear regression. This yields a set of regression coefficients $\beta_i$ for each analog month $i \in [1, N_s]$, which, when linearly combined, reproduce the target pressure field with high fidelity. The same coefficients are subsequently applied to the corresponding target variable anomaly fields from the selected analog months to reconstruct the portion of the target variable anomaly associated with the circulation, providing an estimate of the dynamically driven anomaly.

The random subsampling and reconstruction procedure is repeated $N_r$ times, and the results are averaged to obtain a robust estimate of the dynamic component. The residual component is defined as the difference between the total anomaly and the dynamically reconstructed anomaly. Repeating this procedure for all target months yields timeseries of dynamic and residual precipitation anomalies. The final result is a decomposition of our target variable $X$ into $X = X_{dynamic} + X_{residual}$, from which trends can be calculated.

For extended periods, such as 1980–2100, we maintain consistency by selecting analogs from, and calculating anomalies with respect to, the 1980–2022 climatology (or the first 43 years of the piControl simulation). This ensures that the pool of available analog months remains fixed in size and that analogs are drawn from an approximately stationary climate. Stationarity helps to minimize contamination of the dynamically reconstructed anomaly by non-dynamical changes in the target variable. For example, if precipitation exhibits a strong thermodynamic trend but there is little circulation change over 1980-2100, selecting analog months from the late 21st century would imprint this thermodynamic signal onto the dynamically reconstructed precipitation, artificially enhancing a weak dynamic precipitation trend. Throughout this study, we use $N_a = 50$, $N_s = 30$, and $N_r = 100$, as in \cite{doane2025dynamic}, which correspond to a similar fraction of analog months as employed by \cite{deser_forced_2016}.

\subsection{Conceptual Comparison of Diagnostics}
\label{concepts}
Let us consider the relative advantages of the two methods described above. Starting with the analog method, an immediate advantage is that we can partition P itself into dynamic or thermodynamic contributions, rather than being limited to decomposing P - E as in the moisture budget approach. Moisture budgets may also fail to close in reanalysis products or even in climate models, particularly over regions of strong orography \citep{seager2013diagnostic,tootoonchi_revisiting_2025}. In addition, the analog method does not require reanalysis precipitation, allowing the use of gridded observational datasets \citep{doane2025dynamic}. However, these advantages come with increased methodological uncertainty, arising from the somewhat arbitrary choices of $N_a$, $N_r$, and $N_s$, as well as the definition of the circulation domains used for matching. At the same time, restricting the regression to the nearest $N_a$ analog months allows the method to capture nonlinear relationships despite relying on a linear model: each fit is applied only to a subset of similar circulation states, so the relationship with the target variable is linear within that subset but varies across the full range of states. Repeating this across all months therefore yields a piecewise linear, but overall nonlinear, relationship between SLP and the target variable.

On the other hand, with the analogs we cannot neatly separate the submonthly dynamic variability from monthly variability \citep{branstator1995organization,rennert2009cross}. The fraction of submonthly variability in the $X_{dynamic}$ term is simply the fraction of submonthly variability that is expected from the overall monthly circulation pattern, according to the piecewise linear regression defined above. No two months are perfectly alike in their submonthly statistics, so some of the submonthly dynamic variability of the target field will fall in the $X_{residual}$ term. Indeed, the relationship between submonthly and monthly variability may be different across different regions of circulation space. Therefore, if there exists an overall circulation trend (e.g. towards higher pressure), the residual trend may contain a submonthly dynamic trend if the submonthly dynamics become more strongly/weakly related to the monthly mean circulation. 

Finally, SLP mainly describes the near-surface circulation and may miss any upper-level changes which can still influence precipitation. As a result, one may rewrite the decomposition from the analogs method more explicitly as: 
$X = E[X|\text{SLP}_\text{monthly}] + X_{\text{residual}}$, where the residual explains all physical processes not explained by the monthly-mean SLP pattern.

Turning our attention to the moisture budget, it appears at first that this method gives us more detail than the analogs method, due to the separation into four rather than two terms. The first term, $\bar{\bar{q}}\bar{\mathbf{u}}'$, corresponds to the VIMC anomaly due to monthly mean wind anomalies alone, which is seemingly related to the circulation $E[X|\text{SLP}_\text{monthly}]$ or $X_{dynamic}$ term of the analogs method. Likewise, the $\bar{q}'\,\bar{\bar{\mathbf{u}}}$ term corresponds to the VIMC anomaly due to changes in monthly mean moisture alone. 
However, it is important to note that dynamic as well as thermodynamic processes can lead to changes in the monthly mean moisture field. Advection from arid or humid regions due to the monthly mean circulation pattern will still partly manifest as a change in $\nabla \cdot (\bar{q}'\,\bar{\bar{\mathbf{u}}})$. 
This "mixed dynamic-thermodynamic" process \citep{seager_climate_2019} will be the focus of much of Section \ref{linking}.

The final two terms, $\bar{q}'\,\bar{\mathbf{u}}'$ and $q''\,\mathbf{u}''$, describing monthly and submonthly eddies respectively, represent interactions between moisture and circulation anomalies which inherently combine both thermodynamic and dynamic processes. The submonthly eddy term $q''\,\mathbf{u}''$, capturing synoptic-scale transient contributions to VIMC, cannot be evaluated by the analogs approach which relies on monthly mean data. This ability to explicitly quantify submonthly transient contributions represents a major advantage of the moisture budget framework. In the context of the analog decomposition, both eddy terms could contribute to both $X_{dynamic}$ and $X_{residual}$, since changes in these terms arise from coupled variations in both moisture $q$ and circulation $\mathbf{u}$, making it difficult to attribute them cleanly to either purely dynamic or purely thermodynamic processes.

In summary, while the moisture budget approach uses a robust mathematical framework and offers explicit separation of submonthly transient effects, it is limited by systematic budget closure errors.  Dynamic and thermodynamic contributions also remain stubbornly coupled in several terms. The analog method provides a cleaner separation of circulation-driven change and can use observational or reanalysis precipitation directly, but cannot resolve submonthly dynamics and can conflate different physical processes within its residual term. These complementary strengths and limitations motivate the combined framework we develop in the following sections.

\section{Case Study Application}
\label{casestudy}
To facilitate a direct comparison between the two methods, we apply the analogs approach to VIMC as the target field $X$, rather than to $P-E$ or $P$ alone. While this choice sacrifices one of the analog method's primary advantages, it ensures any difference in results between the analogs and the moisture budget are due to methodological differences alone. We focus our analysis on Central Chile as the primary case study region.

\begin{figure}[ht!]
  \centering
  \includegraphics[width=0.8\textwidth]{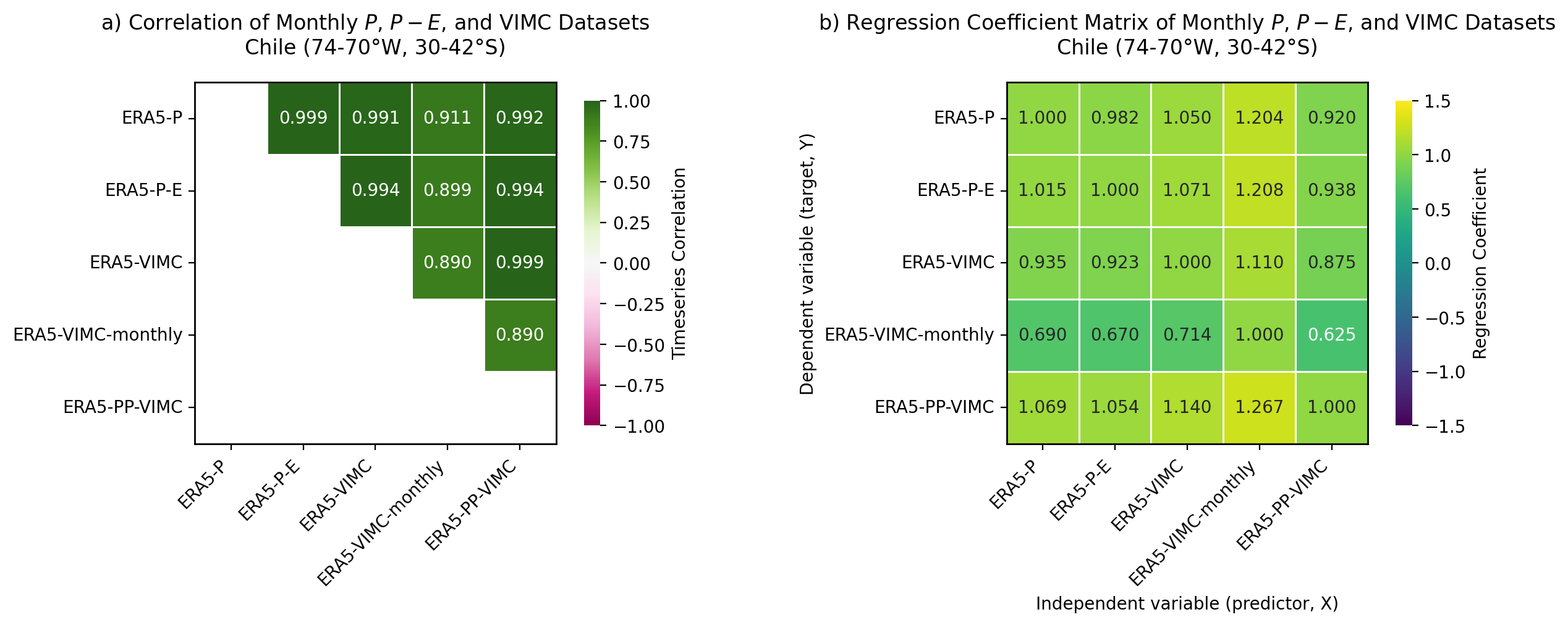}
  \caption{a) Correlation matrix showing monthly timeseries correlations between ERA5 datasets for $P$, $P-E$, VIMC, VIMC-monthly and PP-VIMC over Central Chile (74-70\textdegree W, 30-42\textdegree S). b) Regression coefficient matrix showing regression coefficient of X-variable against Y-variable from the same ERA5 datasets.}
  \label{fig:correl_regress_matrix}
\end{figure}

Figure \ref{fig:correl_regress_matrix} demonstrates that our focus on VIMC directly is justified in Central Chile. First, the directly computed 6-hourly VIMC exhibits strong correlations with both $P$ and $P-E$ ($r > 0.99$), confirming that VIMC is a reliable proxy for precipitation variability in this region. Second, the high correlation and regression coefficient near 1 between $P$ and $P-E$ indicates that evaporation plays a minimal role in wintertime monthly variability (and trend - see Figure \ref{supp:chile_analogs_evaporation}), whereas precipitation is dominant. Third, VIMC and $P-E$ are very well correlated and have a similar magnitude with each other but less so with VIMC-monthly, highlighting the importance of submonthly transient eddies in delivering moisture and generating precipitation in this midlatitude region. Nevertheless, the correlation between VIMC and VIMC-monthly of 0.89 suggests about 80\% of the variance is captured by the monthly mean.
Finally, the close agreement between our directly computed monthly moisture budget and the postprocessed product from \cite{Mayer_ERA5_Budgets_2021} in regression and correlation provides validation of our moisture budget calculations, and supports the use of this framework for diagnosing precipitation changes in Central Chile.

\subsection{Constructed Circulation Analog Decomposition}
\begin{figure}[ht!]
  \centering
  \includegraphics[width=\textwidth]{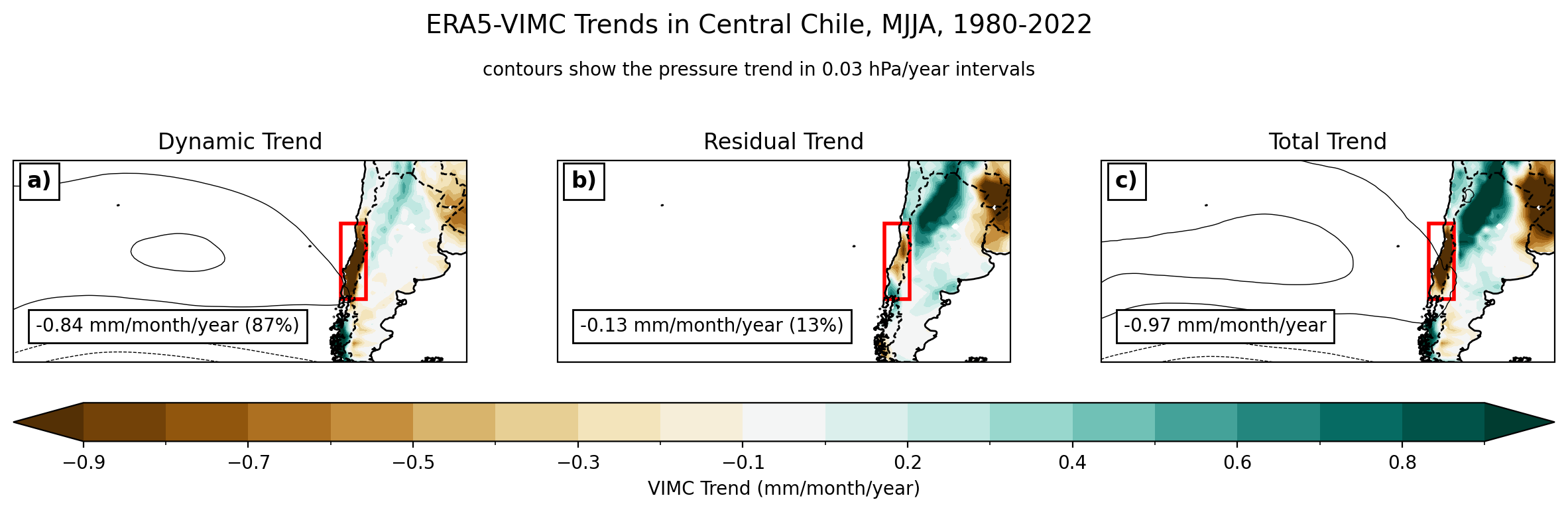}
  \caption{Circulation analog decomposition of VIMC trends in MJJA from 1980-2022 over Central Chile, shown in the red box (74-70\textdegree W, 30-42\textdegree S). Panel a) shows the dynamic trend, b) the residual trend, and c) the total trend. VIMC trends are shown in color and SLP
  trends are shown in contours (0.03hPa/year, solid = positive, dotted = negative).}
  \label{fig:chile_analogs}
\end{figure}

We start by performing a constructed circulation analog decomposition using ERA5 data to find the dynamic and residual trends in wintertime VIMC over the Central Chile region from 1980 to 2022. This is seen in Figure \ref{fig:chile_analogs}, which  shows a total trend of -0.97 mm/month/year, in panel c). We note that this is similar but slightly different to that found in \cite{doane2025dynamic}, as this study uses a shorter period beginning in 1980 as opposed to 1979 and uses reanalysis VIMC rather than observational precipitation datasets. The decomposition shows that the dynamic trend contributes 87\% of the overall trend, and the residual trend forms only 13\%, a similar split to that seen for precipitation in \cite{doane2025dynamic}. The dynamic trend is relatively uniform across the region, and appears to be caused by a trend towards higher pressure across the subtropical SE Pacific. On the other hand, the residual trend is only negative in the central and northern parts of Central Chile. As discussed in section \ref{methods}\ref{concepts}, the residual will contain the thermodynamic trend as well as some proportion of the submonthly dynamic trend.

\subsection{Moisture Budget Decomposition}
We next apply the moisture budget decomposition to the same VIMC trend in Central Chile as the circulation analogs trend. We note the total trend is -0.98 mm/month/year rather than -0.97 mm/month/year as in the analogs: this is simply due to the trend being calculated on the 6-hourly data directly rather than being resampled to monthly frequency as required for the analogs.

\begin{figure}[ht!]
  \centering
  \includegraphics[width=\textwidth]{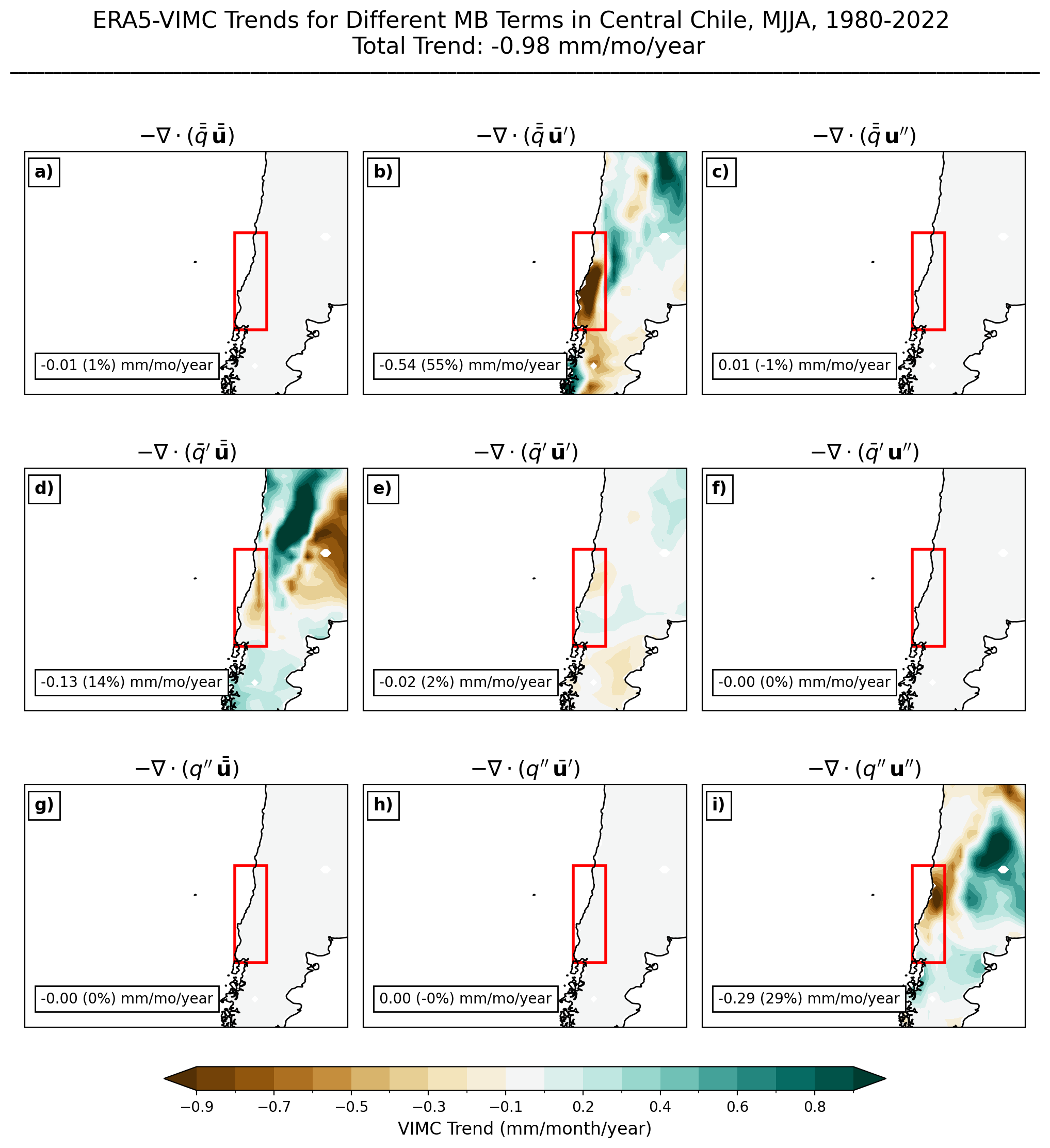}
  \caption{Moisture budget decomposition of VIMC trends in MJJA from 1980-2022 over Central Chile, shown in the red box (74-70\textdegree W, 30-42\textdegree S). The nine panels show the nine terms in the Reynolds decomposition in Eq.~\eqref{expansion}. The top row corresponds to terms with climatological $q$, the middle row those with the monthly anomaly of $q$, and the bottom with the 6-hourly submonthly anomaly of $q$. The left column corresponds to terms with climatological $\mathbf{u}$, the middle column those with the monthly anomaly of $\mathbf{u}$, and the right column with the 6-hourly submonthly anomaly of $\mathbf{u}$.}
  \label{fig:chile_mb}
\end{figure}

In this framework, we find that the trend in moisture convergence due to changes in the monthly mean wind (Figure \ref{fig:chile_mb}b) forms 55\% of the overall trend. This is considerably less than the 87\% in the dynamic trend from the analogs. Changes in monthly mean moisture (Figure \ref{fig:chile_mb}d) contribute 14\% of the convergence trend. The monthly eddies (Figure \ref{fig:chile_mb}e) only contribute 2\%, while the submonthly eddies contribute substantially with 29\% (Figure \ref{fig:chile_mb}i). As expected, the other four terms are negligible due to the relatively small changes in $p_s$.

We note that the dynamic trend in the analogs is approximately equal to the sum of the changes due to the eddies and the monthly mean wind (fully or partially dynamic terms). Additionally, the amplitude and pattern of the residual trend is very similar to the trend in the monthly mean moisture field (often considered a thermodynamic term).  While it appears physically plausible that these terms from both methods correspond to each other, we will show in the following section it is generally not the case.

\section{Linking Moisture Budgets and Analogs}
\label{linking}
In this section, we aim to quantify the relationships between moisture budget terms and constructed circulation analog terms. We increase our sample size by using CESM2-LE data, and analyze both monthly variability and multidecadal trends.

\subsection{Monthly Variability}
\label{monthly_var}
\begin{figure}[ht!]
  \centering
  \includegraphics[width=\textwidth]{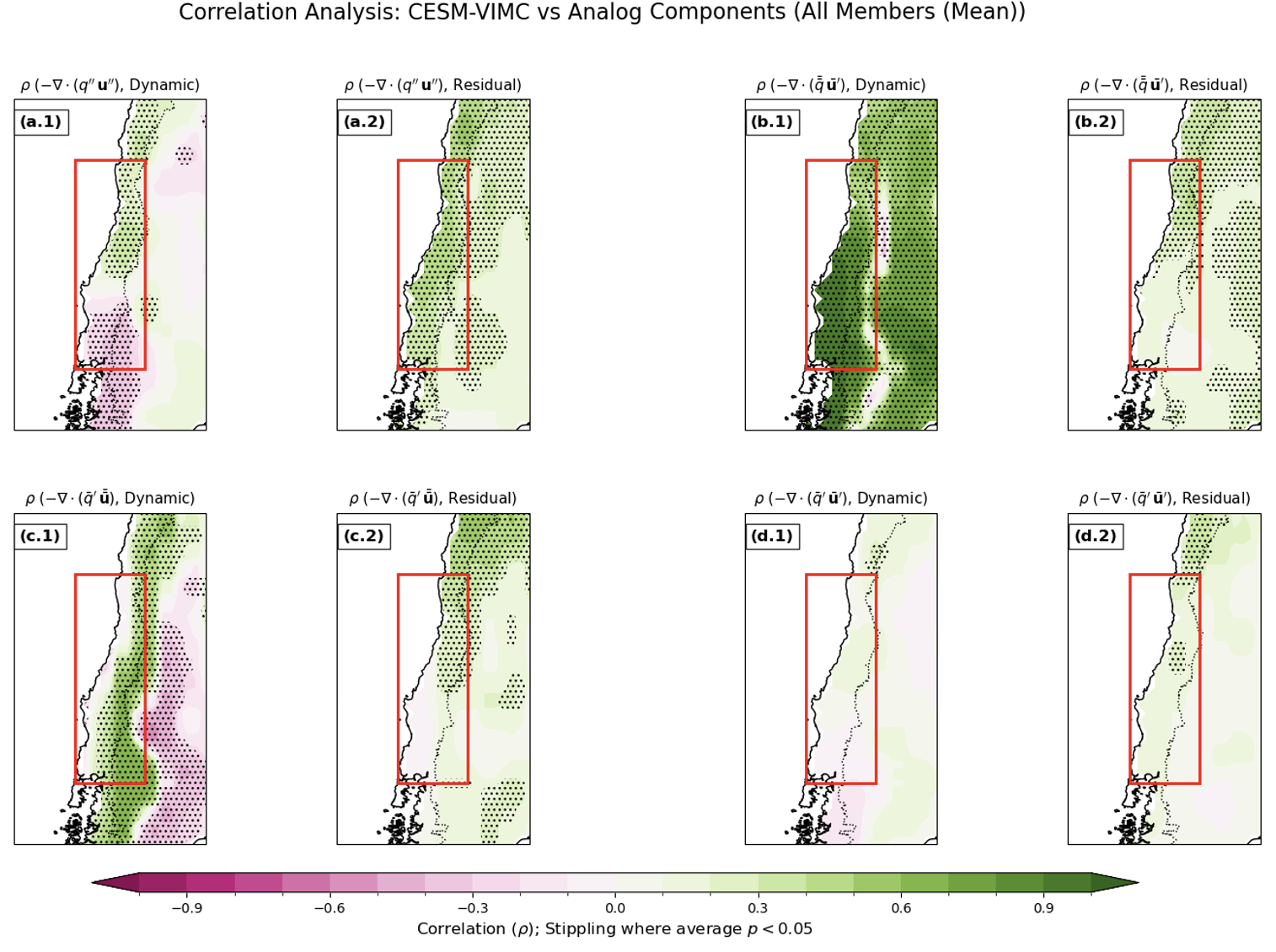}
  \caption{Correlation between moisture budget and analogs components of CESM-VIMC in MJJA from 1980-2022 over Central Chile, shown in the red box (74-70\textdegree W, 30-42\textdegree S). The four panels show the nonzero VIMC terms from the Reynolds decomposition in  Eq.~\eqref{expansion_short}. Subpanels (X.1) show the correlation between the MB term and the dynamic analog term, and subpanels (X.2) show the correlation between the MB term and the residual analog term. Green = positive correlation, pink = negative correlation. Stippling indicates an average 
  $p < 0.05$
  across the 10 ensemble members.}
  \label{fig:cesm10_analogs_mb_variability_both}
\end{figure}

We perform a timeseries correlation analysis between the moisture budget and constructed circulation analog terms, computed across both months and ensemble members, using the 10 CESM2-LE members with submonthly output. We calculate the correlation between the timeseries of each moisture budget term and both the dynamic and residual term from the analogs. We then average across the ensemble members. Figure \ref{fig:cesm10_analogs_mb_variability_both} shows the result; similar patterns hold for ERA5 (Figure \ref{supp:era5_analogs_mb_variability_both}).

Green areas indicate locations with positive correlations between the MB term and the analog term, and pink corresponds to areas with negative correlations between the MB term and the analog term. Figures \ref{fig:cesm10_analogs_mb_variability_both}a.1 and \ref{fig:cesm10_analogs_mb_variability_both}a.2 show that on average over Central Chile, the submonthly transient eddies are weakly positively correlated to both terms. In the southern half of the region, the eddies are in fact slightly negatively correlated with the dynamic term.  The monthly eddies are weakly and generally not significantly correlated to either analog term (Figure \ref{fig:cesm10_analogs_mb_variability_both}d). Both results are somewhat expected, due to the nature of these terms involving covarying changes in both moisture $q$ and wind $\mathbf{u}$. Furthermore, Figure \ref{fig:cesm10_analogs_mb_variability_both}b shows a much higher correlation between the dynamic analog term and the monthly wind anomaly moisture budget term than the residual analog term.

More interestingly, we see in Figure \ref{fig:cesm10_analogs_mb_variability_both}c that the monthly moisture anomaly term is also generally more highly correlated with the dynamic term than the residual term, especially in the interior of the region. Therefore, this "mixed dynamic-thermodynamic" moisture budget term appears significantly more driven by dynamics rather than truly thermodynamic changes on timescales of monthly variability.
It is not \textit{a priori} surprising that dynamics dominate monthly variability of the $-\nabla \cdot (\bar{q}'\,\bar{\bar{\mathbf{u}}})$ term, given that a forced thermodynamic trend may take many decades to emerge from internal variability. However, it is not clear that the same will hold true for multidecadal trends in the same term, as we show in the next subsections.

\subsection{Trends over the Satellite Era}
\label{satellite-trends}
In this section, we use data from all 100 members of CESM2-LE as well as the CAM6 piControl simulation described in Section \ref{methods}a. We also use both the ERA5 reanalysis as above and also the JRA-3Q reanalysis. Note that we must use VIMC-monthly for both CESM2-LE and the piControl (and for consistency, the reanalyses), as only monthly output is available. There are therefore no submonthly transient eddies in any of the results in this section, however, Figure \ref{fig:correl_regress_matrix} shows that ERA5-VIMC-monthly has high correlation with ERA5-VIMC, which is also true for the other datasets (not shown).

\begin{figure}[ht!]
  \centering
  \includegraphics[width=\textwidth]{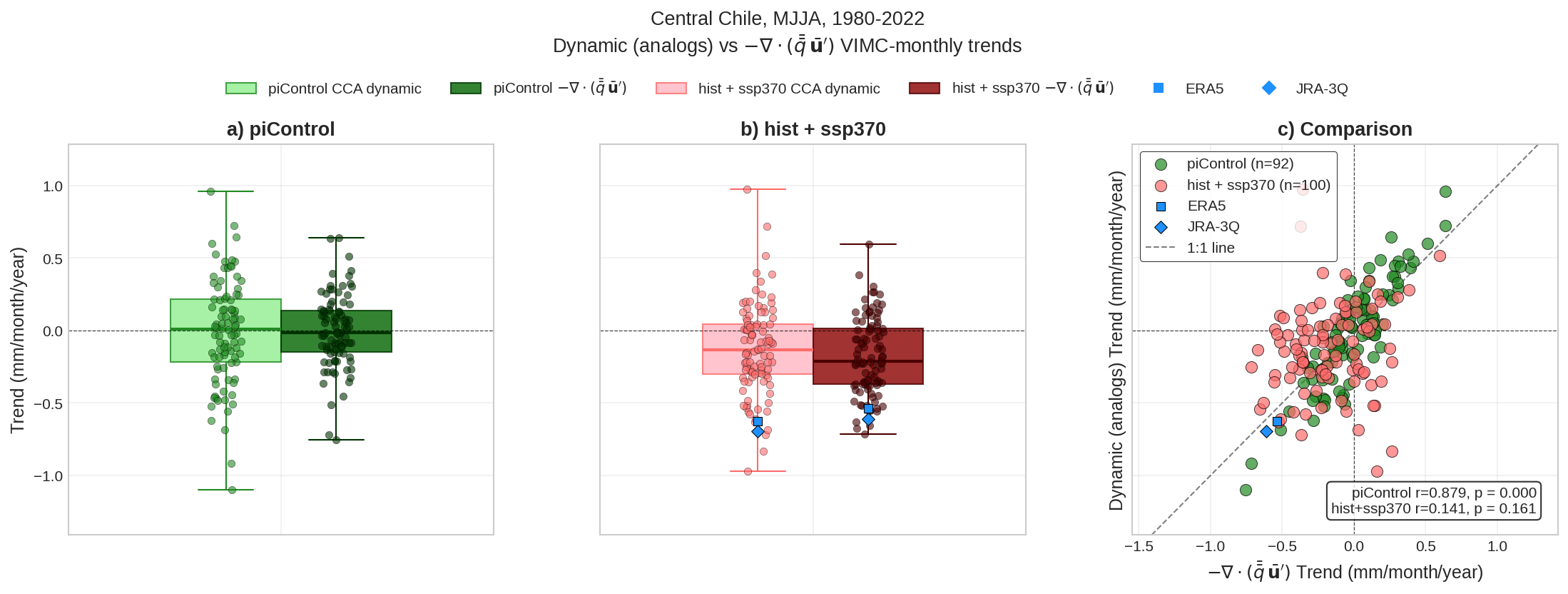}
  \caption{'Dynamic' trends of VIMC-monthly in piControl and CESM2-LE (hist + ssp370) in Central Chile. a) Boxplot of dynamic components of 42-year trends in piControl: light green = dynamic (analogs), dark green = $-\nabla \cdot (\bar{\bar{q}}\,\bar{\mathbf{u}}')$. b) Boxplot of dynamic components of trends from 1980-2022 in CESM2-LE: light red = dynamic (analogs), dark red = $-\nabla \cdot (\bar{\bar{q}}\,\bar{\mathbf{u}}')$. c) Scatterplot of analogs vs MB trends for CESM2-LE and piControl. The lower right box shows correlations and p-values. Blue squares show corresponding ERA5 trends, blue diamonds show JRA-3Q trends.}
  \label{fig:picontrol_dynamic_Chile_2022}
\end{figure}

We start by analyzing the 'dynamic' trends of VIMC across CESM2-LE (hist + ssp370 scenarios) in Central Chile from 1980 to 2022 in Figure \ref{fig:picontrol_dynamic_Chile_2022}b. For the moisture budget, dynamic trends are those due to trends in the monthly mean $\mathbf{u}$ coupled with climatological $q$, i.e. $-\nabla \cdot \left(\bar{\bar{q}}\,\bar{\mathbf{u}}'\right)$ and for the analogs these are changes in VIMC expected from the regional SLP pattern, $E[-\nabla \cdot (q\mathbf{u})|\text{SLP}_\text{monthly}]$. We see that both the analog and MB methods give a large spread in the estimated dynamic trends, and that the mean trend from the MB method is larger. Both methods show a significant negative ensemble mean dynamic trend ($p \ll 0.01$), as expected, but trends of this magnitude readily occur in the piControl experiment (Figure \ref{fig:picontrol_dynamic_Chile_2022}a) over the same length time period. In the synthetic piControl ensemble, the dynamic trends from both methods appear correlated (Figure \ref{fig:picontrol_dynamic_Chile_2022}c). Perhaps surprisingly, this does not appear to be true in the forced CESM2-LE. Indeed, the average difference between the decomposition methodologies is about 0.1 mm/month/year: this is the same order of magnitude as the mean dynamic trend according to both the MB and the analogs. The observed trends are consistent between ERA5 and JRA-3Q and towards the bottom of the CESM2-LE and piControl spread. The fact that both methods give very similar estimates for both ERA5 and JRA-3Q appears coincidental given the spread of differences between the methods found in the large ensemble.

\begin{figure}[ht!]
  \centering
  \includegraphics[width=\textwidth]{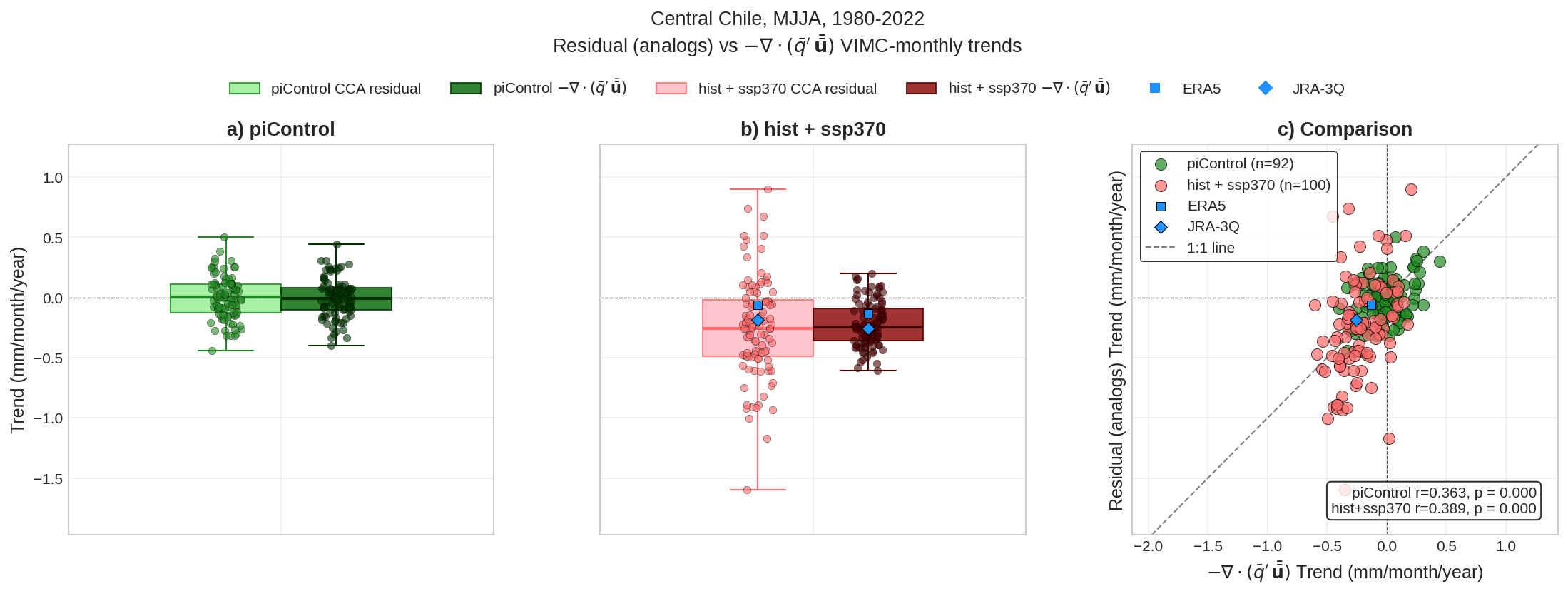}
  \caption{'Thermodynamic' trends of VIMC-monthly in piControl and CESM2-LE (hist + ssp370) in Central Chile. a) Boxplot of thermodynamic components of 42-year trends in piControl: light green = thermodynamic (analogs), dark green = $-\nabla \cdot (\bar{q}'\,\bar{\bar{\mathbf{u}}})$. b) Boxplot of thermodynamic components of trends from 1980-2022 in CESM2-LE: light red = thermodynamic (analogs), dark red = $-\nabla \cdot (\bar{q}'\,\bar{\bar{\mathbf{u}}})$. c) Scatterplot of analogs vs MB trends for CESM2-LE and piControl. The lower right box shows correlations and p-values. Blue squares show corresponding ERA5 trends, blue diamonds show JRA-3Q trends.}
  \label{fig:picontrol_thermodynamic_Chile_2022}
\end{figure}

Turning our attention to the 'thermodynamic' trends (Figure \ref{fig:picontrol_thermodynamic_Chile_2022}), we see similar patterns. Both the analog and MB methods reveal comparable (and significant) mean thermodynamic trends across the ensembles, though the spread is much larger for the analogs. As with the dynamic trends, thermodynamic trends of similar magnitude to those in CESM2-LE and the reanalyses are easily simulated in piControl. 

Individual ensemble members again show considerable differences between the two methods (Figure \ref{fig:picontrol_thermodynamic_Chile_2022}c), with discrepancies of similar magnitude to the mean trends themselves. Why is there such a discrepancy between the methods? It cannot be due to submonthly variability in this case, as we are using VIMC-monthly for both the analogs and the moisture budget. The coupled moisture budget term $-\nabla \cdot (\bar{q}'\,\bar{\mathbf{u}}')$ is also quite small in this region. While the residual analog and $-\nabla \cdot (\bar{q}'\,\bar{\bar{\mathbf{u}}})$ trends are significantly correlated ($p < 0.05$) as expected,we also find that the \textit{dynamic} analog trends are also significantly correlated with $-\nabla \cdot (\bar{q}'\,\bar{\bar{\mathbf{u}}})$ (Supplementary Figure \ref{supp:piControl_dynamic_analogs_thermodynamic_mb_Chile_2022}). This suggests that the dynamical influences on monthly moisture variability also hold for multidecadal trends (Section \ref{linking}\ref{monthly_var}), as expanded on in Section \ref{linking}\ref{120-year-trends}.

Regardless of methodological choice, both the dynamic and thermodynamic observed VIMC-monthly trends in Central Chile over the satellite era remain within the range of preindustrial variability, despite the CESM2-LE ensemble mean showing a small forced trend.

\subsection{120-year Trends}
\label{120-year-trends}
\begin{figure}[ht!]
  \centering
  \includegraphics[width=\textwidth]{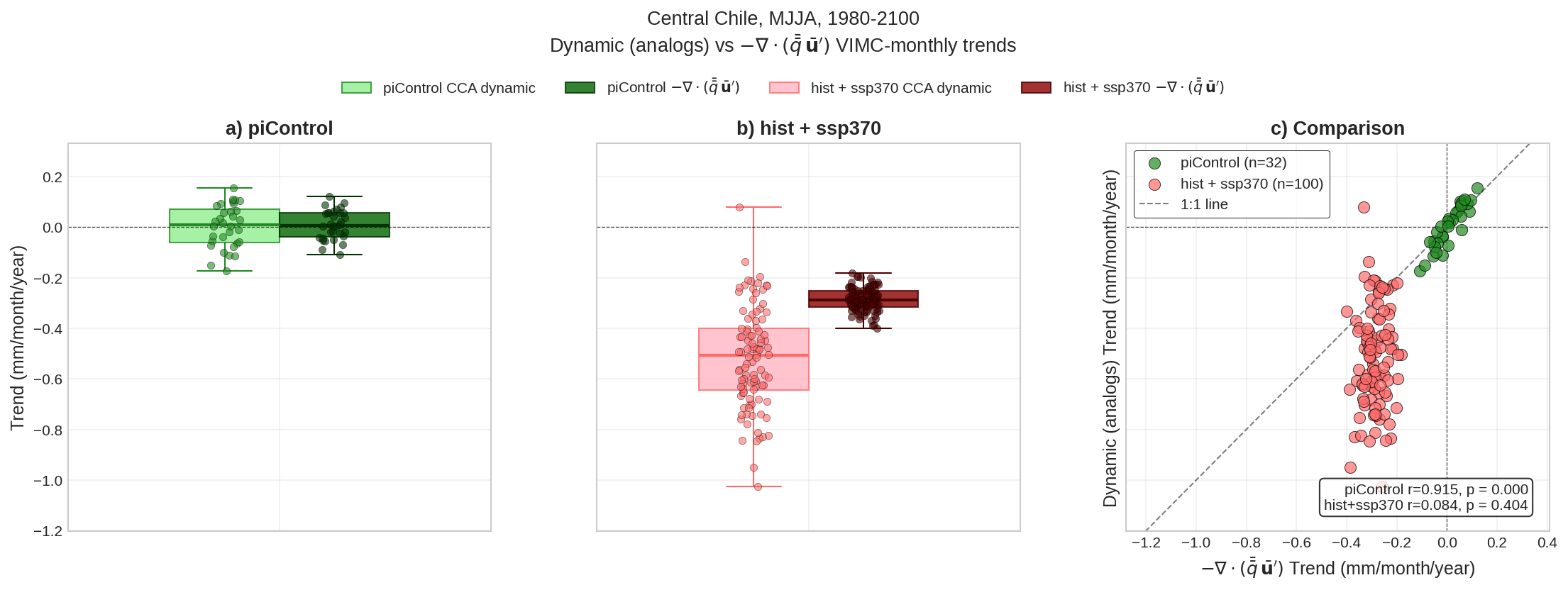}
  \caption{'Dynamic' trends of VIMC-monthly in piControl and CESM2-LE (hist + ssp370) in Central Chile. a) Boxplot of dynamic components of 120-year trends in piControl: light green = dynamic (analogs), dark green = $-\nabla \cdot (\bar{\bar{q}}\,\bar{\mathbf{u}}')$. b) Boxplot of dynamic components of trends from 1980-2100 in CESM2-LE: light red = dynamic (analogs), dark red = $-\nabla \cdot (\bar{\bar{q}}\,\bar{\mathbf{u}}')$. c) Scatterplot of analogs vs MB trends for CESM2-LE and piControl. The lower right box shows correlations and p-values.}
  \label{fig:picontrol_dynamic_Chile_2100}
\end{figure}

This subsection is structured similarly to Section \ref{linking}\ref{satellite-trends}. We start by analyzing the dynamic trends in CESM2-LE from 1980 to 2100 (Figure \ref{fig:picontrol_dynamic_Chile_2100}b). Both methods give a strong dynamic drying trend, but the mean trend  differs very significantly between the two methods. The moisture budget trends have much lower variance across the ensembles and have a mean trend of approximately -0.3 mm/month/year, whereas the analogs have much larger variance but a mean trend of -0.5 mm/month/year. The reason for this difference in spread is unclear (see Section \ref{discussion} for further discussion). Over this longer period, all members are outside the range of the corresponding term in piControl (Figure \ref{fig:picontrol_dynamic_Chile_2100}a) for the moisture budget, and almost all are for the analogs. Both methods again appear well correlated in piControl but not in the CESM2-LE.

\begin{figure}[ht!]
  \centering
  \includegraphics[width=\textwidth]{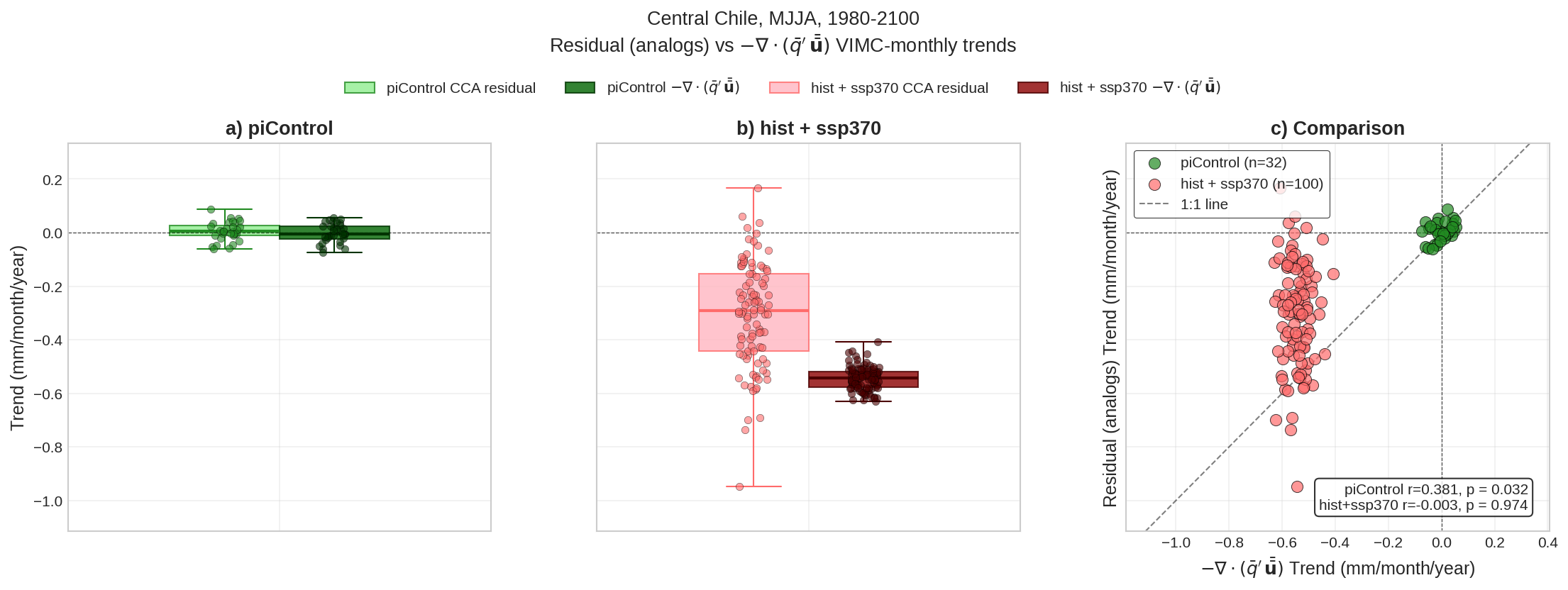}
  \caption{'Thermodynamic' trends of VIMC-monthly in piControl and CESM2-LE (hist + ssp370) in Central Chile. a) Boxplot of thermodynamic components of 120-year trends in piControl: light green = thermodynamic (analogs), dark green = $-\nabla \cdot (\bar{q}'\,\bar{\bar{\mathbf{u}}})$. b) Boxplot of thermodynamic components of trends from 1980-2100 in CESM2-LE: light red = thermodynamic (analogs), dark red = $-\nabla \cdot (\bar{q}'\,\bar{\bar{\mathbf{u}}})$. c) Scatterplot of analogs vs MB trends for CESM2-LE and piControl. The lower right box shows correlations and p-values.}
  \label{fig:picontrol_thermodynamic_Chile_2100}
\end{figure}

Turning our attention to the 'thermodynamic' trends for the same period in Figure \ref{fig:picontrol_thermodynamic_Chile_2100}, we see again that the variance is much larger in the analogs. The reason for this is unclear. However, the mean trend is much lower in the analogs than the moisture budget. The thermodynamic trends are well outside the piControl distribution in the moisture budget, but this is not the case for the analogs. 

We additionally note that there is now no significant correlation in trends between the two 'thermodynamic' methods applied to CESM2-LE over this longer time period. However, in Supplementary Figure \ref{supp:piControl_dynamic_analogs_thermodynamic_mb_Chile_2100}, we find that there is still a significant ($p < 0.05$) correlation between the dynamic analog trends and the trends in  $-\nabla \cdot (\bar{q}'\,\bar{\bar{\mathbf{u}}})$. This suggests that in the presence of such a strong dynamic trend, the 'thermodynamic' trends in $-\nabla \cdot (\bar{q}'\,\bar{\bar{\mathbf{u}}})$ are heavily contaminated. This also explains why the dynamic analogs trend in Figure \ref{fig:picontrol_dynamic_Chile_2100}b is larger: the circulation trends that lead to an advection of less water vapour fall in the dynamic analogs term but in the $-\nabla \cdot (\bar{q}'\,\bar{\bar{\mathbf{u}}})$ in the moisture budget. Therefore, in regions of strong dynamic change, one must be cautious about interpretations of thermodynamic change in moisture budget decompositions, since the dynamics can play an important role in moisture trends.

\section{Re-evaluating Thermodynamic Trends Using a Combined Framework}
\begin{figure}[ht!]
  \centering
  \includegraphics[width=\textwidth]{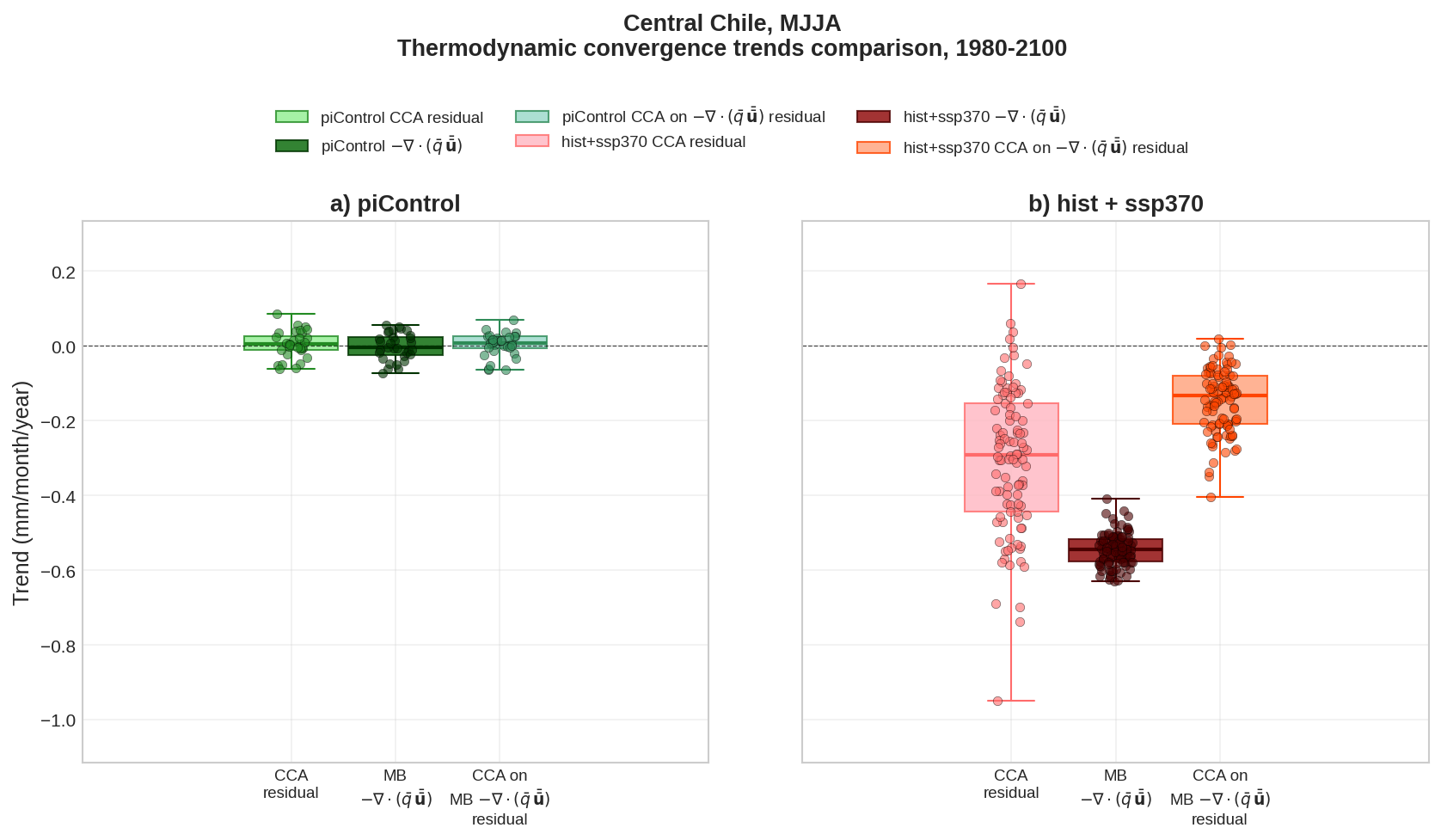}
  \caption{'Thermodynamic' trends of VIMC-monthly in piControl and CESM2-LE (hist + ssp370) using analogs, MB, and the combined framework in Central Chile. a) Boxplot of thermodynamic components of 120-year trends in piControl: light green = analogs, dark green = $-\nabla \cdot (\bar{q}'\,\bar{\bar{\mathbf{u}}})$, pale green = combined. b) Boxplot of thermodynamic components of trends from 1980-2100 in CESM2-LE: light red = analogs, dark red = $-\nabla \cdot (\bar{q}'\,\bar{\bar{\mathbf{u}}})$, orange = combined.}
  \label{fig:thermodynamic_comparison_Chile_2100}
\end{figure}

In this section, we combine both the moisture budget and analogs approach to generate a more constrained estimate of purely thermodynamic change in Central Chile. To do so, we apply the constructed circulation analogs directly to the $-\nabla \cdot (\bar{q}'\,\bar{\bar{\mathbf{u}}})$ term of the moisture budget. This separates out the humidity change due to dynamics into the dynamic term of the analogs, leaving only the the truly thermodynamic term behind in the residual, i.e. the change in $q$ that is unrelated to changes in the circulation. The choice of $-\nabla \cdot (\bar{q}'\,\bar{\bar{\mathbf{u}}})$ also removes any submonthly variability. We first show an example of this in Figure \ref{fig:thermodynamic_comparison_Chile_2100}.

In Figure \ref{fig:thermodynamic_comparison_Chile_2100}b, we see a boxplot of trends for the moisture budget, analog, and combined approaches applied to CESM2-LE data from 1980 to 2100. The MB and analog data is the same as in Figure \ref{fig:picontrol_thermodynamic_Chile_2100}b. We note that the combined data shows a much weaker thermodynamic trend than both the MB and analog residual approaches. Similar to the analogs, a large majority but not all of the members show a drying trend. In Figure \ref{fig:thermodynamic_comparison_Chile_2100}a we show the same for piControl data. We note that the distribution of trends in the CESM2-LE (hist + ssp370 scenario) is largely but not completely outside the distribution from the piControl runs for the combined approach, in contrast to the moisture budget.

\begin{figure}[ht!]
  \centering
  \includegraphics[width=\textwidth]{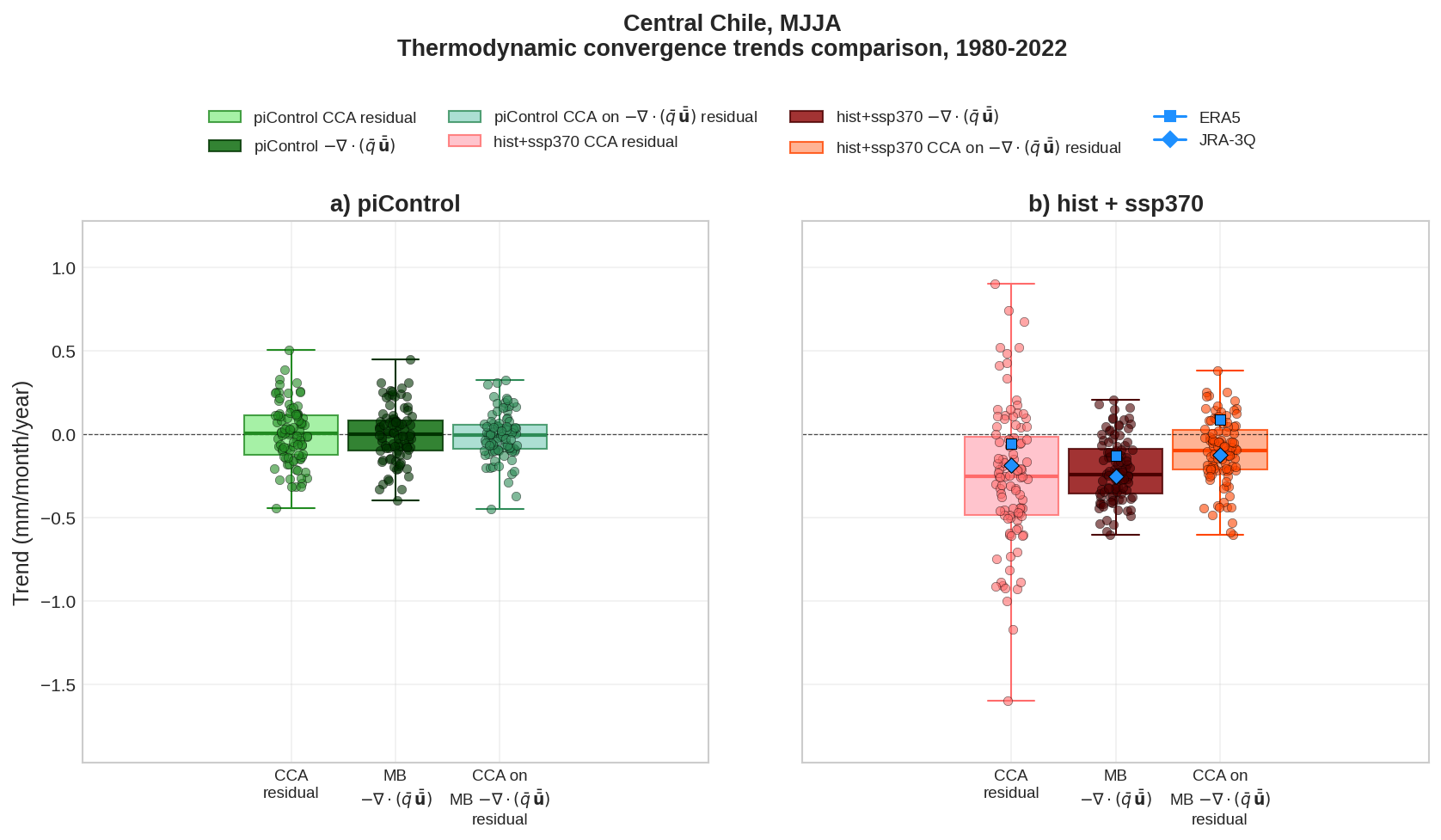}
  \caption{'Thermodynamic' trends of VIMC-monthly in piControl and CESM2-LE (hist + ssp370) using analogs, MB, and the combined framework in Central Chile. a) Boxplot of thermodynamic components of 42-year trends in piControl: light green = analogs, dark green = $-\nabla \cdot (\bar{q}'\,\bar{\bar{\mathbf{u}}})$, pale green = combined. b) Boxplot of thermodynamic components of trends from 1980-2022 in CESM2-LE: light red = analogs, dark red = $-\nabla \cdot (\bar{q}'\,\bar{\bar{\mathbf{u}}})$, orange = combined. Blue squares show ERA5 trends, blue diamonds show JRA-3Q trends.}
  \label{fig:thermodynamic_comparison_Chile_2022}
\end{figure}

We now return to the shorter trend over the satellite era in Central Chile using this framework. In Figure \ref{fig:thermodynamic_comparison_Chile_2022}, we see that over the the 42 year period from 1980 to 2022, there is significant overlap with the equivalent length periods in piControl in all of the analogs, moisture budget and combined methods. The combined method reveals that the thermodynamic trend in CESM2-LE varies across ensemble members, with some showing positive trends and others negative trends. The ERA5 thermodynamic trend, which is weakly negative using either of the two methods alone, becomes positive when using the combined framework, and the trend in JRA-3Q also weakens (blue points in Figure \ref{fig:thermodynamic_comparison_Chile_2022}). In summary, there is no evidence for a thermodynamic trend in observed monthly moisture convergence over the satellite era in Central Chile , whereas in CESM2, there is evidence for a weak trend in the ensemble mean. 

\begin{figure}[ht!]
  \centering
  \includegraphics[width=\textwidth]{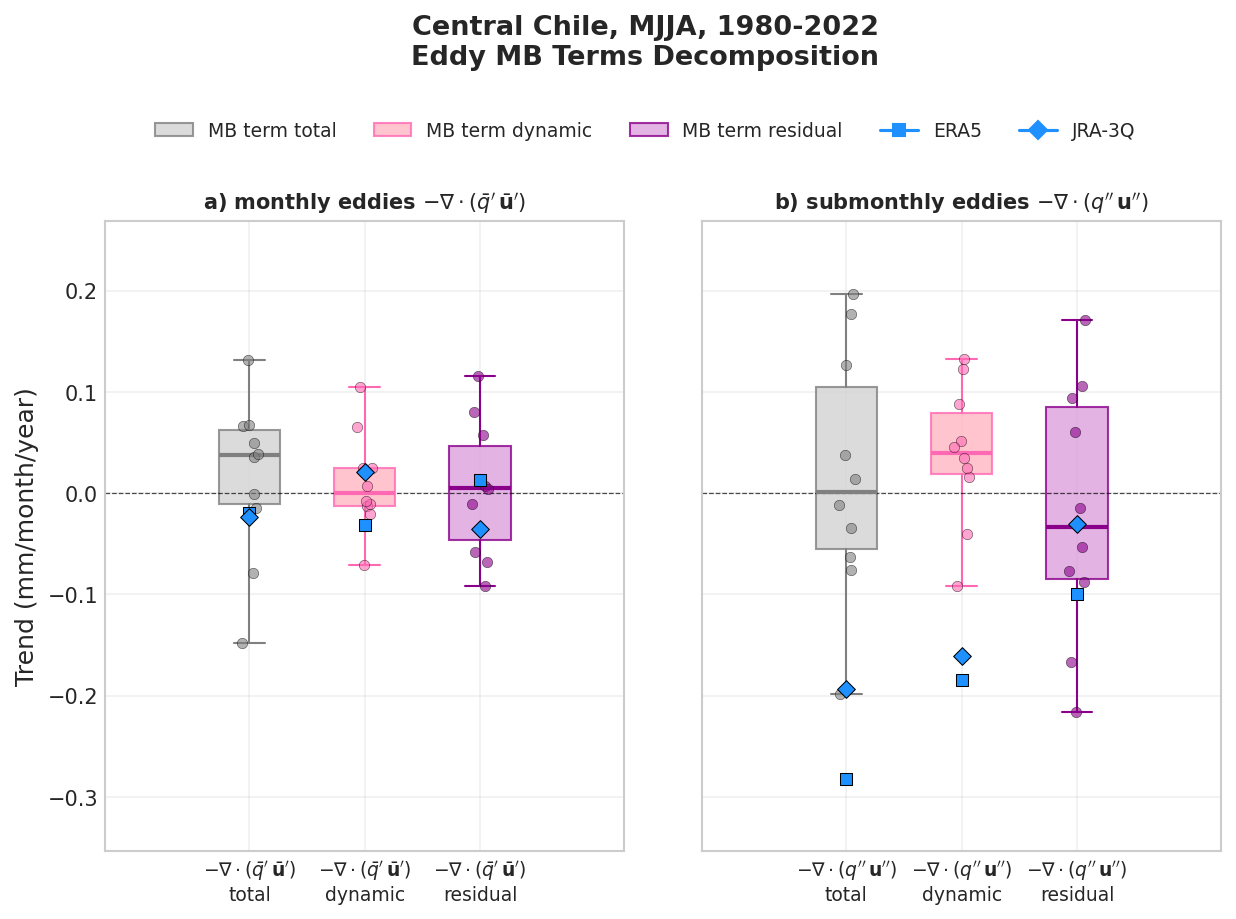}
  \caption{Analogs decomposition of VIMC eddy trends in the 10 members of CESM2-LE (hist + ssp370), ERA5 and JRA-3Q from 1980-2022. a) shows monthly eddy $-\nabla \cdot (\bar{q}'\,\bar{\mathbf{u}}')$ trends, b) shows submonthly eddy $-\nabla \cdot (q''\mathbf{u}'')$ trends. Gray shows the total trend, pink the dynamic component, red the residual component. Blue squares show ERA5 trends, blue diamonds show JRA-3Q trends.}
  \label{fig:eddy_terms_Chile_2022}
\end{figure}

Finally, we can use the combined framework developed above to decompose the monthly and submonthly eddy terms. For comparability between the two, we use the 10 members of CESM2-LE with 6-hourly data. Unfortunately, we cannot compare this to preindustrial variability due to the lack of 6-hourly piControl data. We see in Figure \ref{fig:eddy_terms_Chile_2022}a that there is a very weak overall wetting trend for the monthly eddies in total over the satellite era: it is approximately zero in both reanalyses and CESM2-LE. The analogs decomposition shows that there are no strong dynamic tendencies in monthly convergence either. However, this is not the case for the submonthly eddies as shown in Figure \ref{fig:eddy_terms_Chile_2022}b. CESM2-LE simulates a mean dynamic convergence trend in the submonthly eddies, and a weak mean residual divergence trend. Intriguingly, both reanalyses show a strong dynamic divergence trend outside the 10 ensemble members.

\section{Conclusions and Discussion}
\label{discussion}
In this study, we introduce a refined framework for isolating thermodynamic trends in vertically integrated moisture convergence (VIMC) using a moisture budget approach together with constructed circulation analogs. We have applied this framework to Central Chile, a region experiencing severe drying trends in observations \cite{boisier_anthropogenic_2018, garreaud_central_2020,
doane2025dynamic}, and compared it to a standard moisture budget decomposition and dynamical adjustment approach. Our key findings are as follows:

1) Estimates of the relative contribution of dynamic and thermodynamic processes to VIMC trends are highly method-dependent. The analogs method and the moisture budget method agree on the mean sign and approximate magnitude of the dynamic trend, though they differ in their quantitative estimates and can even differ in sign for some ensemble members. Over both the 42- and 120-year periods, estimates of dynamic and thermodynamic trends using each method are uncorrelated across the CESM2-LE. However, trends in CESM2-LE over the satellite era are largely within the range of preindustrial variability regardless of the method used.

2) We find that traditional moisture budget decompositions can significantly overestimate purely thermodynamic trends in regions of strong circulation trends. The $-\nabla \cdot (\bar{q}'\,\bar{\bar{\mathbf{u}}})$ term contains a substantial contribution from dynamic changes when circulation anomalies are large. This is particularly evident in the 120-year trends, where the correlation between the analogs' dynamic term and $-\nabla \cdot (\bar{q}'\,\bar{\bar{\mathbf{u}}})$ trends remains significant while the correlation with the analogs' residual term is not. The influence of sea level pressure changes on this term is further demonstrated by the significant correlations between SLP and $-\nabla \cdot (\bar{q}'\,\bar{\bar{\mathbf{u}}})$ trends over 120-year periods shown in Figure \ref{supp:thermodynamic_mb_on_psl}. Therefore this "dynamic-thermodynamic" term \citep{seager_climate_2019} can be majority dynamic in regions experiencing large circulation changes, such as Mediterranean-type climates subject to the poleward expansion of the subtropical dry zones.

3) By combining the analogs and moisture budget approaches, we can obtain a more refined estimate of purely thermodynamic change. This combined framework reveals that the true thermodynamic trend in Central Chile is weaker than suggested by either method alone. Over the satellite era, there is little evidence that thermodynamic trends in VIMC have emerged from preindustrial variability in either CESM2-LE or reanalysis datasets. 

\begin{figure}[ht!]
  \centering
  \includegraphics[width=1\textwidth]{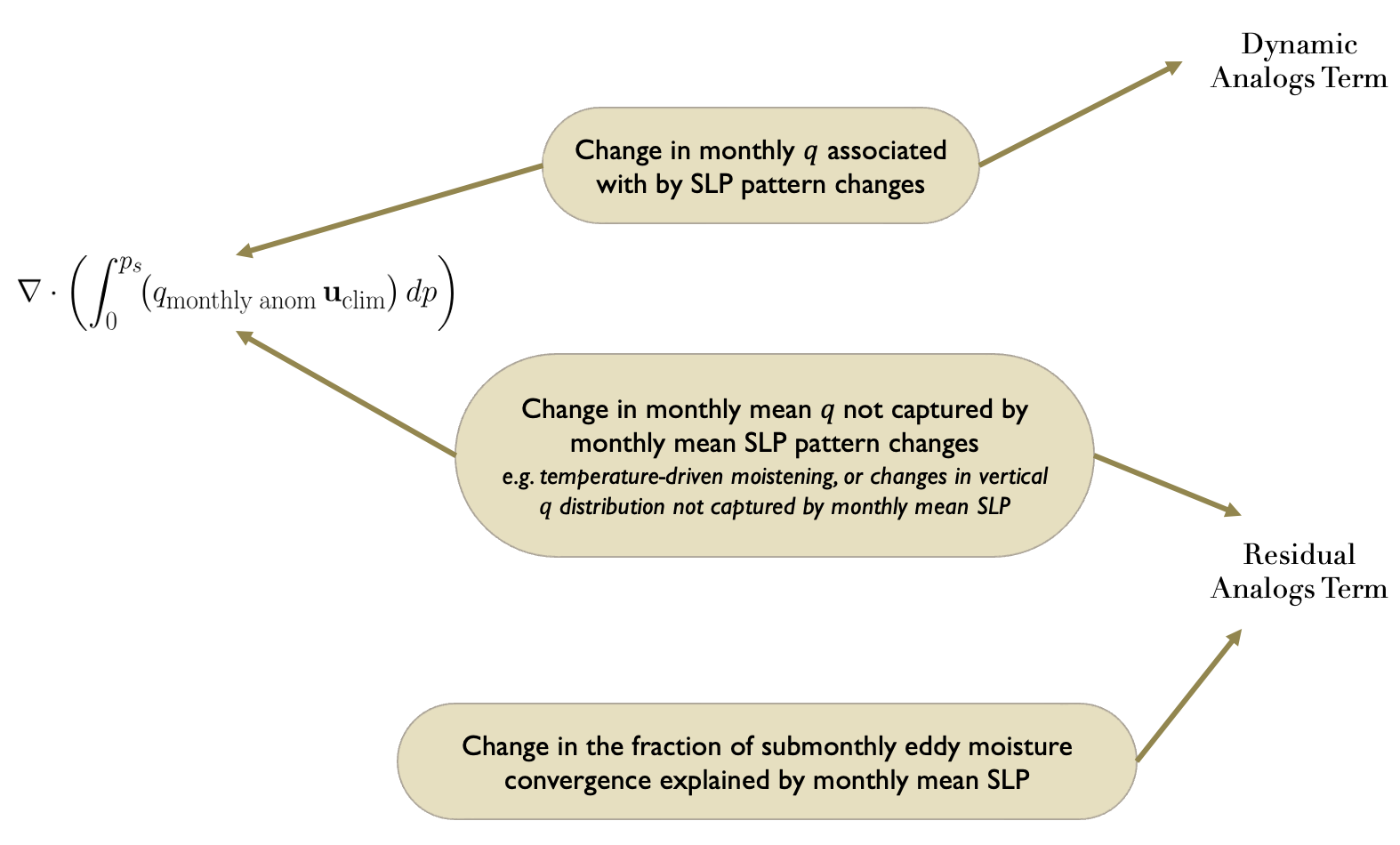}
  \caption{Schematic summary of how moisture budget and circulation analog estimates of circulation-independent thermodynamic trends are contaminated by other effects. Arrows point from a physical process to the term which contains it in the moisture budget and circulation analogs.}
  \label{fig:thermodynamic_schematic}
\end{figure}

We can also use this approach to measure dynamic and thermodynamic contributions to changes in submonthly eddies, although several caveats and limitations should be noted. Firstly, the residual trend will only be equal to the thermodynamic trend in the submonthly eddies if the relationship between monthly SLP and submonthly variability remains stationary. It also assumes monthly SLP explains all submonthly dynamic variability (a flawed assumption, as discussed in Section \ref{methods}\ref{concepts}). A schematic summary of the methods' approach to `thermodynamic' trends is shown in Figure \ref{fig:thermodynamic_schematic}.

However, the large spread across the ensemble members with the analog method compared to the moisture budget method (even over a 120-year time period) remains concerning. One plausible explanation is instability in the regression coefficients, $\beta_i$, as discussed by \cite{pfleiderer_evaluation_2026}. Because the analog predictors are deliberately highly collinear, the coefficients can become ill-conditioned, large, and sensitive to small changes in the selected analog set. On the other hand, the resampling of the analog selection $N_r$ times should reduce the sensitivity to any single analog set and yield more stable coefficients. To test whether this coefficient instability was a factor, we repeated the analysis using cross-validated ridge regression, which explicitly penalizes large coefficients. The resulting dynamic and residual trends were identical to those calculated using OLS regression, therefore we believe coefficient instability is unlikely to be a primary factor. In addition, we find no evidence that analog quality degrades over the 120-year period. The method continues to produce high-fidelity reconstructions using analogs drawn from the 1980–2022 period throughout the simulation, through 2100.

We attempted to improve the constructed circulation analogs method by conditioning the analogs on monthly means of submonthly metrics, such as 6-hourly variance, in addition to SLP to better capture submonthly variability. However, we found no significant improvement in the proportion of submonthly eddy variance explained by the dynamic analogs term. This suggests that the monthly mean SLP analogs already capture approximately the maximum amount of submonthly variance that can be explained by a monthly mean field in these regions. As a result, our combined approach provides an alternative to the robust but computationally expensive percentile-matching method of \cite{siler2023diagnosing} for evaluating dynamic and thermodynamic submonthly contributions.

The focus on VIMC trends means we do not directly address changes in precipitation, which can also depend on evaporation and moisture storage terms. In Central Chile VIMC variability is closely tied to precipitation and $P-E$ variability (Figure \ref{fig:correl_regress_matrix}), but this is not necessarily true in other regions, which may limit the validity of moisture budgets for diagnosing causes of precipitation change. On the other hand, the analogs method inherently has some statistical error and requires sufficient data from an approximately stationary distribution to construct robust analogs. This limits its application to shorter time periods, and exact numerical values may vary due to sampling uncertainty.

While this study finds consistent relationships across two reanalysis products and the state-of-the-art CESM2 model, future research using a multi-model ensemble would enhance the robustness of our conclusions. Expanding our analysis to other regions, especially ones without strong dynamic trends, would also provide a stronger foundation. We briefly explore this for South Australia in Supplementary Figures \ref{supp:piControl_dynamic_SAus_2100} and \ref{supp:thermodynamic_comparison_SAus_2100}. Similar to Central Chile, there is no significant correlation between the MB and analog methods across the ensemble. However, here 
the 120-year dynamic trends in this region are weak compared to preindustrial variability (Supplementary Figure \ref{supp:piControl_dynamic_SAus_2100}). As a result, the mean thermodynamic MB trend is similar before or after the analogs are applied (Supplementary Figure \ref{supp:thermodynamic_comparison_SAus_2100}), unlike in Chile. However, in this region the convergence trend from monthly eddies $-\nabla \cdot (\bar{q}'\,\bar{\mathbf{u}}')$ contributes significantly (not shown), explaining the large difference between the residual of total VIMC and the $-\nabla \cdot (\bar{q}'\,\bar{\bar{\mathbf{u}}})$ term here. We also note the similar spread in both the analogs and moisture budget method in this region, implying the difference in spread over Central Chile may not be representative of these methods more generally. 

Indeed, the contrast between these two regions underscores the necessity of evaluating hydroclimate trends through multiple diagnostic lenses. By placing moisture budget decompositions and constructed circulation analogs on an equal footing, we highlight that these distinct approaches both provide useful perspectives on the causes of hydroclimate change. Constructed circulation analogs are best placed to show the role of monthly mean circulation change, including any nonlinearities, but the moisture budget provides a more detailed physical description of the processes at hand. Our results demonstrate that discrepancies between these methods are not merely errors, but signatures of the relative importance of moisture, circulation and submonthly processes. The parallel use of both frameworks can offer a more robust way to understand the full drivers of regional hydroclimate trends. In Central Chile, the high degree of dynamic-thermodynamic drying suggests that the region's recent negative precipitation trend is even more dependent on South Pacific circulation anomalies than a standard moisture budget might suggest. 

\clearpage
\acknowledgments
This study was supported the UK Natural Environment Research Council (NERC) Award NE/S007474/1.

%
%
\datastatement
ECMWF Reanalysis v5 (ERA5) data used in this study are available at \cite{era5pressurelevels} and \cite{hersbach2023era5}. JRA-3Q data used in this study is available at \cite{ncar_gdex_dataset_d640000}. The Community Earth System Model version 2 (CESM) Large Ensemble data are available at \url{https://esgf-node.llnl.gov/projects/cmip6/}.


%






%



\bibliographystyle{ametsocV6}
\bibliography{references}

\clearpage
\begin{center}
\textbf{\large Supplementary Material}
\end{center}
\setcounter{equation}{0}
\setcounter{figure}{0}
\setcounter{table}{0}
\setcounter{page}{1}
\makeatletter
\renewcommand{\theequation}{S\arabic{equation}}
\renewcommand{\thefigure}{S\arabic{figure}}
\renewcommand{\bibnumfmt}[1]{[S#1]}
\renewcommand{\citenumfont}[1]{S#1}

\renewcommand{\thefigure}{S\arabic{figure}}
\renewcommand{\thetable}{S\arabic{table}}
\renewcommand{\theequation}{S\arabic{equation}}
\setcounter{figure}{0}
\setcounter{table}{0}
\setcounter{equation}{0}

\label{SI}

\begin{figure}[ht!]
  \centering
  \includegraphics[width=\textwidth]{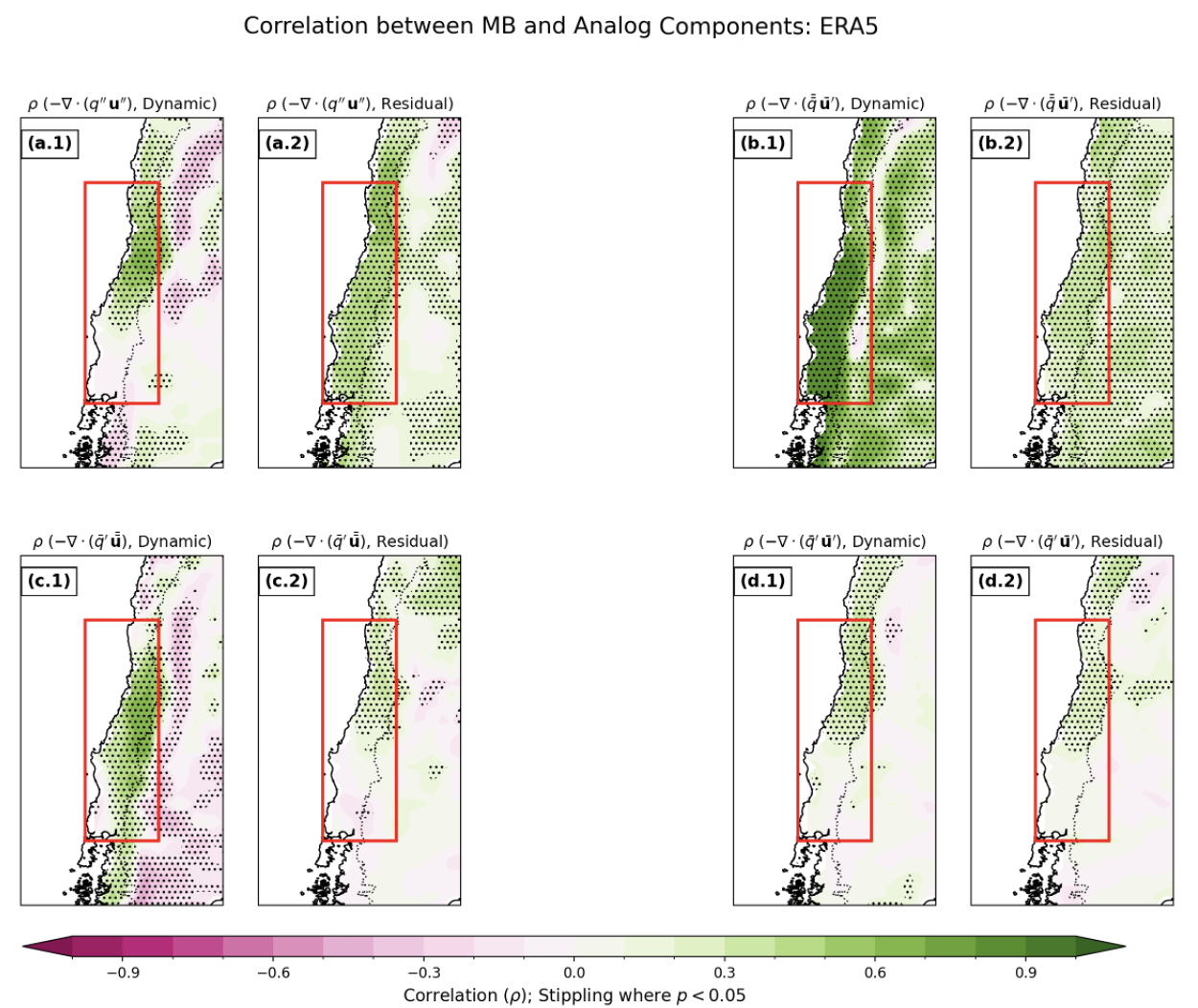}
  \caption{Correlation between moisture budget and analogs components of ERA5-VIMC in MJJA from 1980-2022 over Central Chile, shown in the red box (74-70\textdegree W, 30-42\textdegree S). The four panels show the nonzero VIMC terms from the Reynolds decomposition. Subpanels (X.1) show the correlation between the MB term and the dynamic analog term, and subpanels (X.2) show the correlation between the MB term and the residual analog term. Green = positive correlation, pink = negative correlation. Stippling indicates 
  $p < 0.05$
  .}
  \label{supp:era5_analogs_mb_variability_both}
\end{figure}

\begin{figure}[ht!]
  \centering
  \includegraphics[width=\textwidth]{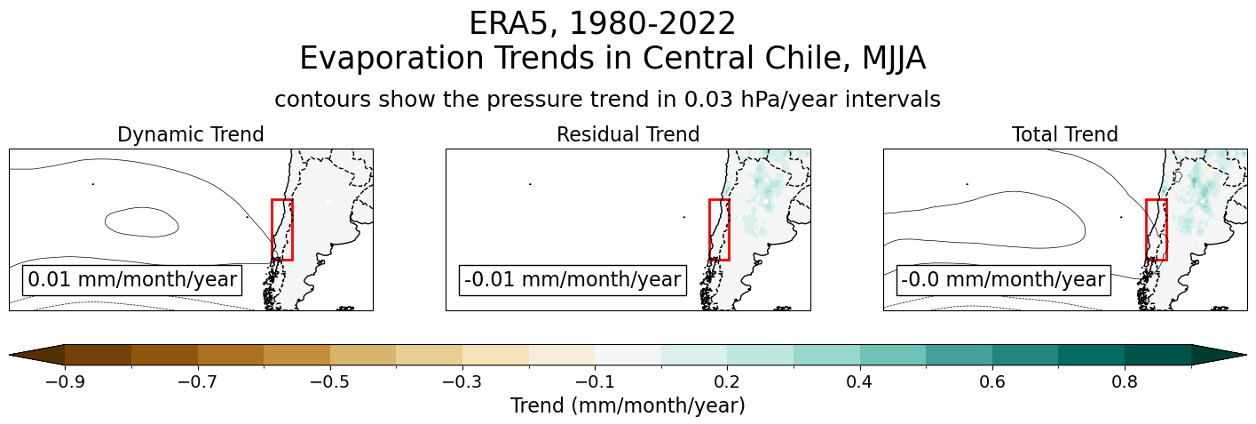}
  \caption{Circulation analog decomposition of ERA5 evaporation trends in MJJA from 1980-2022 over Central Chile, shown in the red box (74-70\textdegree W, 30-42\textdegree S). Panel a) shows the dynamic trend, b) the residual trend, and c) the total trend. Trends are shown in color and SLP
  trends are shown in contours (0.03hPa/year, solid = positive, dotted = negative).}
  \label{supp:chile_analogs_evaporation}
\end{figure}

\begin{figure}[ht]
  \centering
  \includegraphics[width=\textwidth]{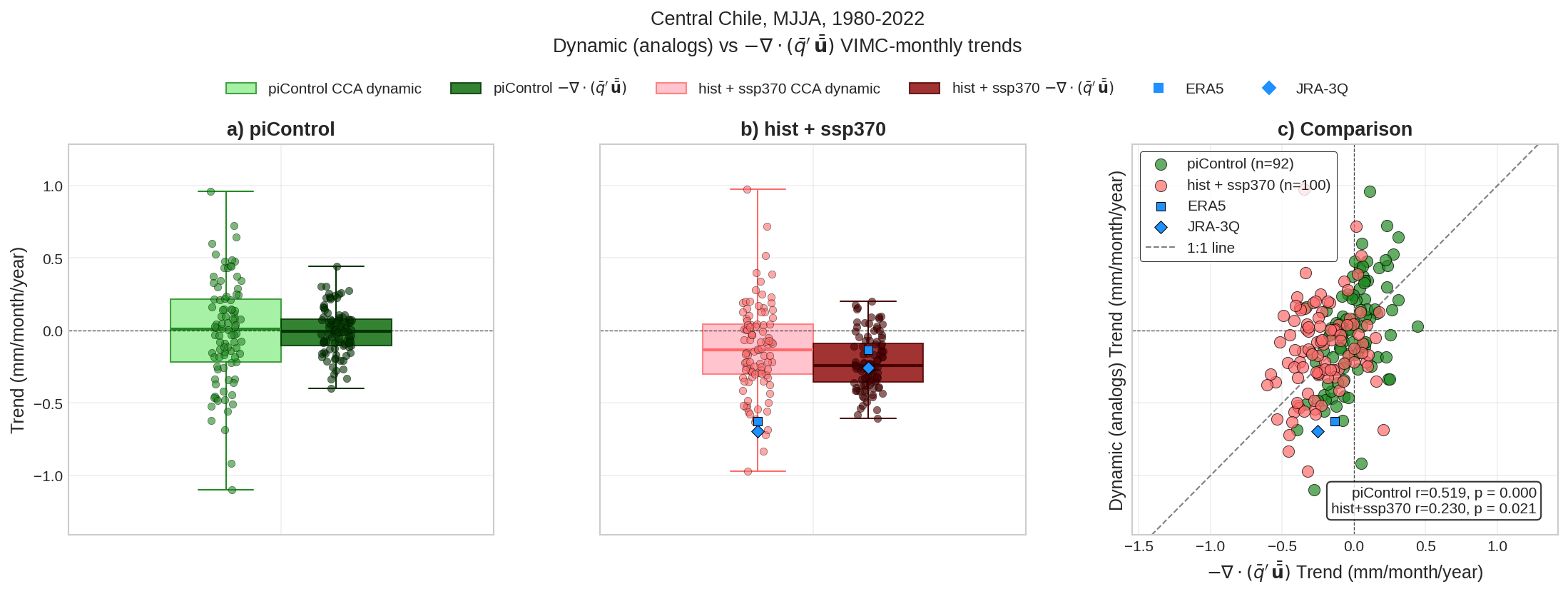}
  \caption{'Dynamic' and 'thermodynamic' trends in piControl and CESM2-LE (hist + ssp370) in Central Chile. a) Boxplot of dynamic components of 42-year trends in piControl: light green = dynamic (analogs), dark green = $-\nabla \cdot (\bar{q}'\,\bar{\bar{\mathbf{u}}})$. b) Boxplot of dynamic components of trends from 1980-2022 in CESM2-LE: light red = dynamic (analogs), dark red = $-\nabla \cdot (\bar{q}'\,\bar{\bar{\mathbf{u}}})$. c) Scatterplot of analogs vs MB trends for CESM2-LE and piControl. The lower right box shows correlations and p-values. Blue squares show corresponding ERA5 trends, blue diamonds show JRA-3Q trends.}
  \label{supp:piControl_dynamic_analogs_thermodynamic_mb_Chile_2022}
\end{figure}

\begin{figure}[ht]
  \centering
  \includegraphics[width=\textwidth]{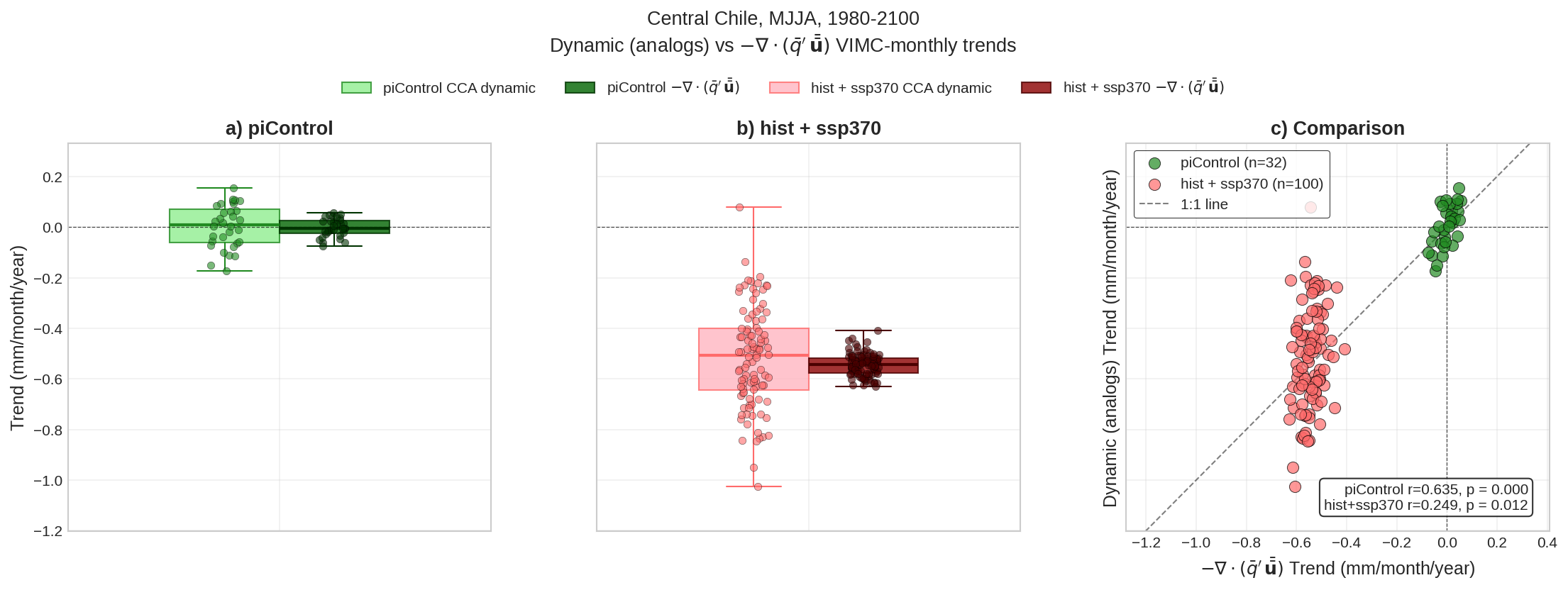}
  \caption{'Dynamic' and 'thermodynamic' trends in piControl and CESM2-LE (hist + ssp370) in Central Chile. a) Boxplot of dynamic components of 120-year trends in piControl: light green = dynamic (analogs), dark green = $-\nabla \cdot (\bar{q}'\,\bar{\bar{\mathbf{u}}})$. b) Boxplot of dynamic components of trends from 1980-2100 in CESM2-LE: light red = dynamic (analogs), dark red = $-\nabla \cdot (\bar{q}'\,\bar{\bar{\mathbf{u}}})$. c) Scatterplot of analogs vs MB trends for CESM2-LE and piControl. The lower right box shows correlations and p-values.}
  \label{supp:piControl_dynamic_analogs_thermodynamic_mb_Chile_2100}
\end{figure}

\begin{figure}[ht]
  \centering
  \includegraphics[width=\textwidth]{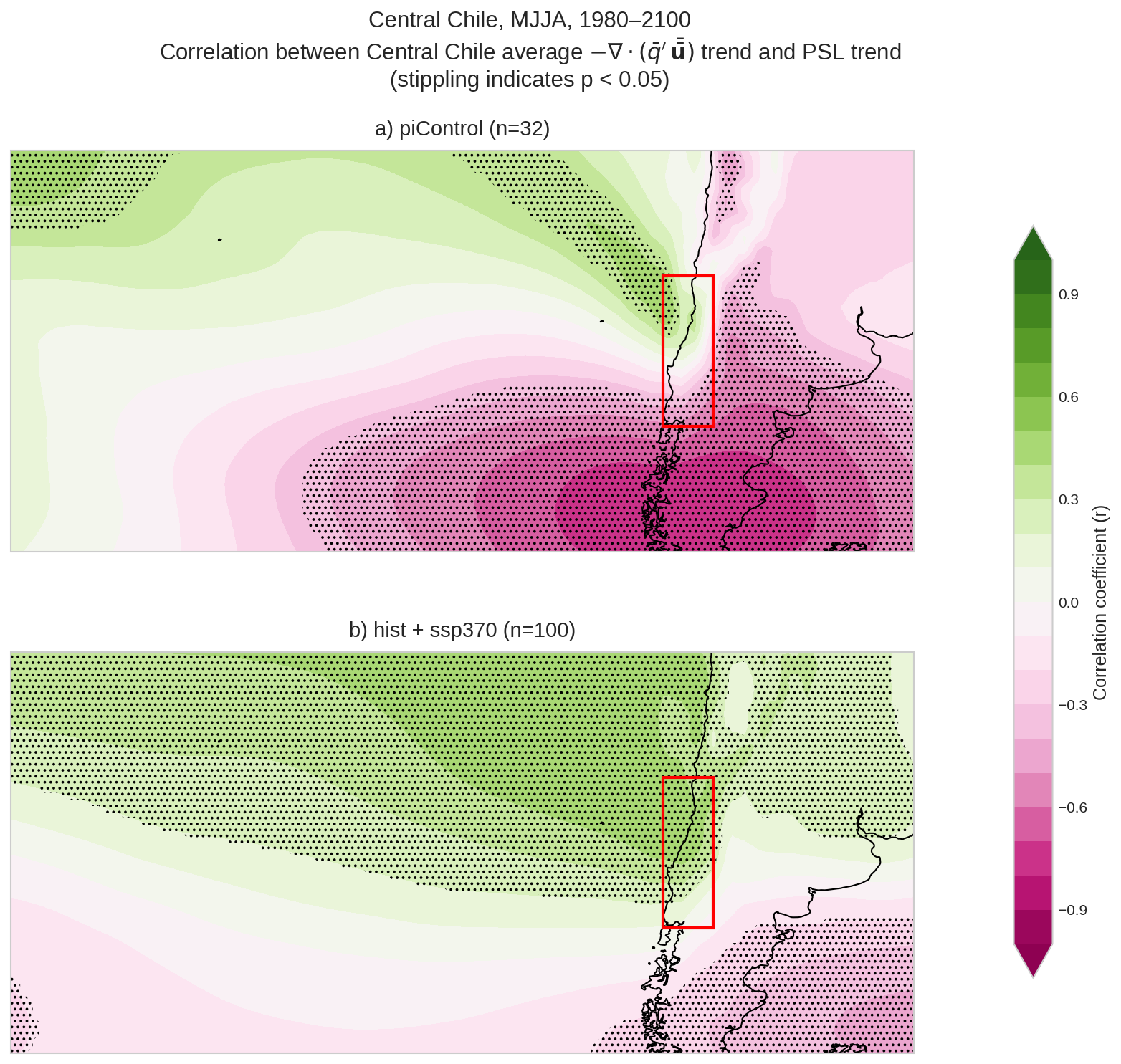}
  \caption{Correlation between PSL trends and the area-averaged $-\nabla \cdot (\bar{q}'\,\bar{\bar{\mathbf{u}}})$ trends at each gridpoint, across all  ensemble members. Green shows positive correlations, pink negative, and stippling indicates an average $p < 0.05$. a) shows the results from piControl, and b) shows the results from CESM2-LE (hist + ssp370).}
  \label{supp:thermodynamic_mb_on_psl}
\end{figure}

\begin{figure}[ht]
  \centering
  \includegraphics[width=\textwidth]{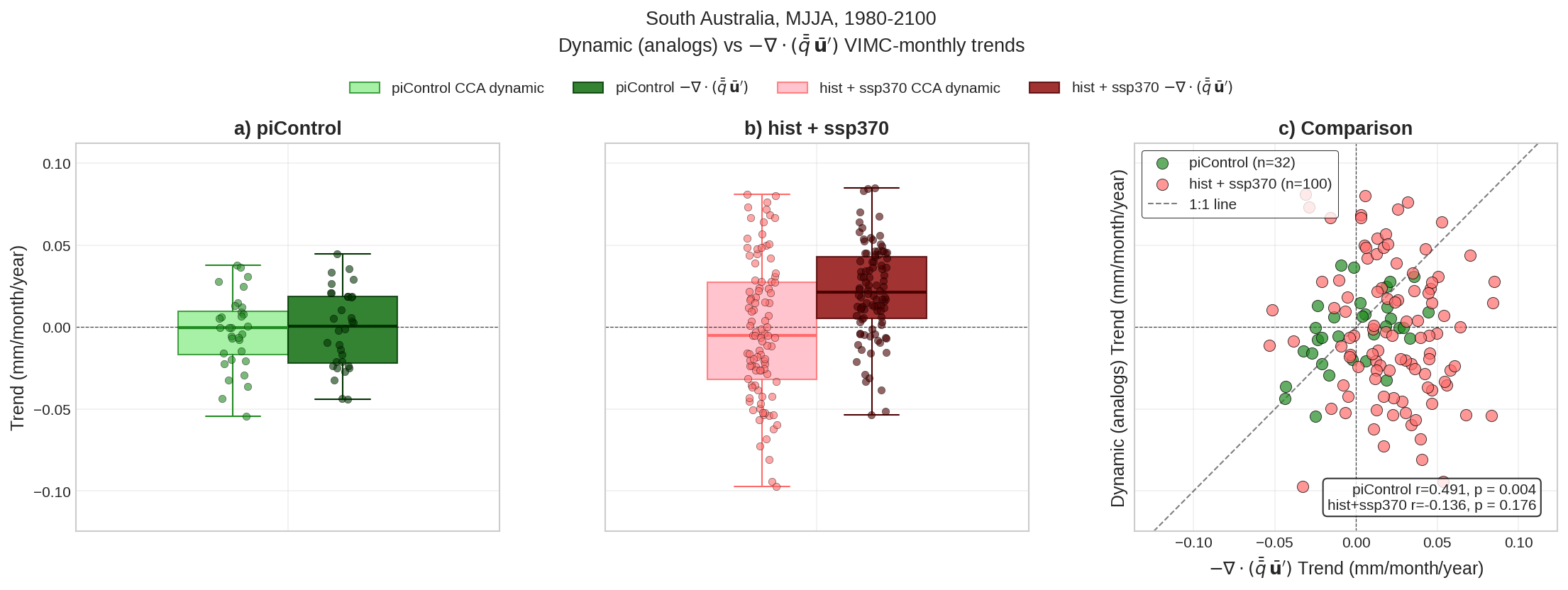}
  \caption{'Dynamic' and 'thermodynamic' trends in piControl and CESM2-LE (hist + ssp370) in South Australia. a) Boxplot of dynamic components of 120-year trends in piControl: light green = dynamic (analogs), dark green = $-\nabla \cdot (\bar{q}'\,\bar{\bar{\mathbf{u}}})$. b) Boxplot of dynamic components of trends from 1980-2100 in CESM2-LE: light red = dynamic (analogs), dark red = $-\nabla \cdot (\bar{q}'\,\bar{\bar{\mathbf{u}}})$. c) Scatterplot of analogs vs MB trends for CESM2-LE and piControl. The lower right box shows correlations and p-values.}
  \label{supp:piControl_dynamic_SAus_2100}
\end{figure}

\begin{figure}[ht!]
  \centering
  \includegraphics[width=\textwidth]{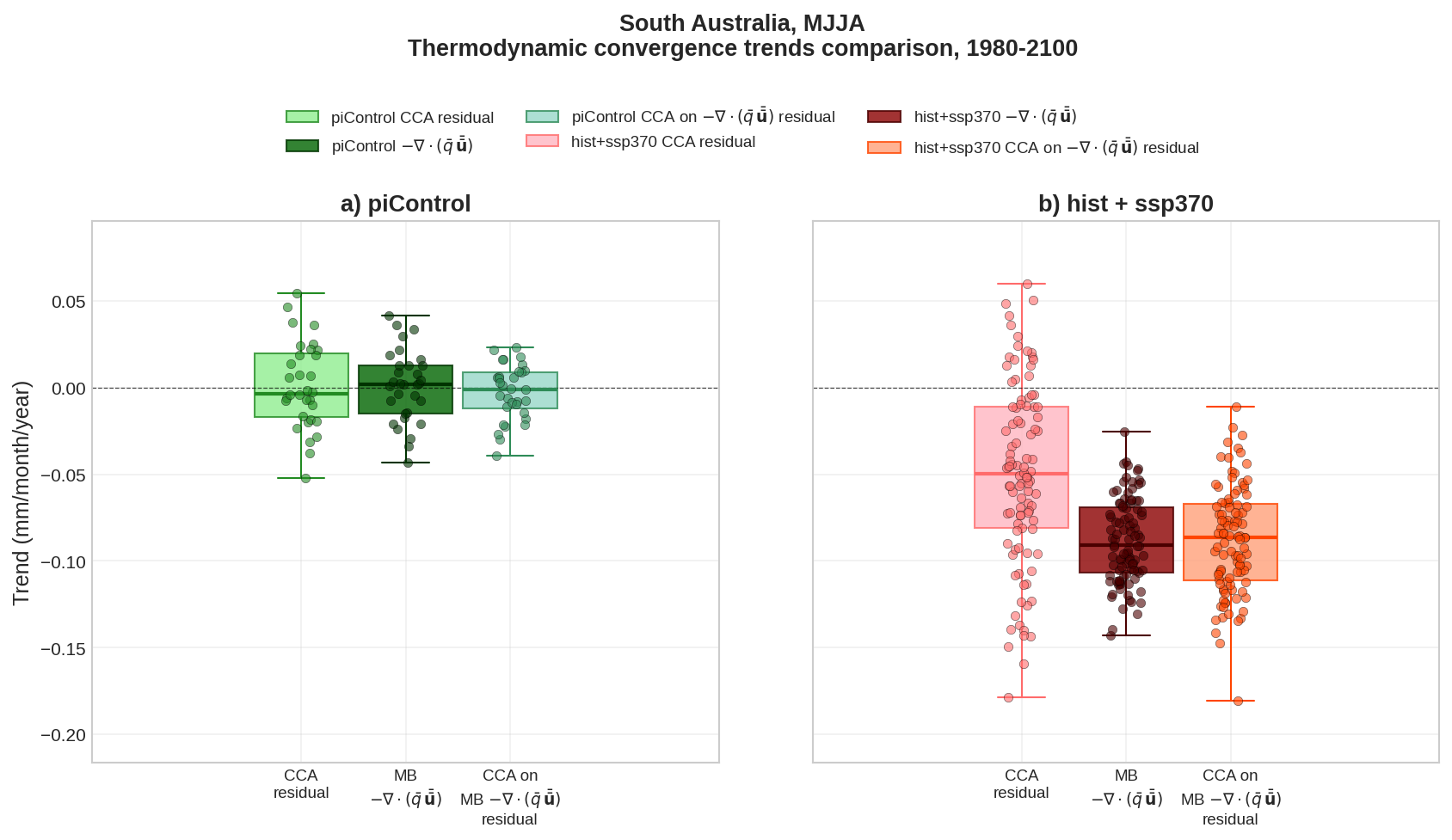}
  \caption{'Thermodynamic' trends in piControl and CESM2-LE (hist + ssp370) using analogs, MB, and the combined framework in South Australia. a) Boxplot of thermodynamic components of 120-year trends in piControl: light green = analogs, dark green = $-\nabla \cdot (\bar{q}'\,\bar{\bar{\mathbf{u}}})$, pale green = combined. b) Boxplot of thermodynamic components of trends from 1980-2100 in CESM2-LE: light red = analogs, dark red = $-\nabla \cdot (\bar{q}'\,\bar{\bar{\mathbf{u}}})$, orange = combined.}
  \label{supp:thermodynamic_comparison_SAus_2100}
\end{figure}

\end{document}